\documentclass[aps,twocolumn,superscriptaddress]{revtex4-1}

\usepackage{breqn}
\usepackage{graphicx}
\usepackage{dcolumn}   % needed for some tables
\usepackage{bm}        % for math
\usepackage{amssymb}   % for math
\usepackage{soul}
\usepackage{gensymb}
\usepackage{enumerate}
\usepackage{multirow}

\usepackage[breaklinks=true,colorlinks=true,linkcolor=blue,urlcolor=blue,citecolor=blue]{hyperref}

\usepackage{float}

\begin{document}

\title{How accurate are flux-tube (local) gyrokinetic codes in modeling energetic particle effects on core turbulence?}

\author{A.~Di Siena} 
\affiliation{Max Planck Institute for Plasma Physics Garching 85748 Germany}
\author{T.~Hayward-Schneider} 
\affiliation{Max Planck Institute for Plasma Physics Garching 85748 Germany}
\author{P.~Mantica} 
\affiliation{Istituto per la Scienza e Tecnologia Dei Plasmi Consiglio Nazionale Delle Ricerche Milano Italy}
\author{J.~Citrin} 
\affiliation{DIFFER—Dutch Institute for Fundamental Energy Research Eindhoven The Netherlands}
\author{F.~Vannini} 
\affiliation{Max Planck Institute for Plasma Physics Garching 85748 Germany}
\author{A.~Bottino} 
\affiliation{Max Planck Institute for Plasma Physics Garching 85748 Germany}
\author{T.~G\"orler}
\affiliation{Max Planck Institute for Plasma Physics Garching 85748 Germany}
\author{E.~Poli} 
\affiliation{Max Planck Institute for Plasma Physics Garching 85748 Germany}
\author{R.~Bilato} 
\affiliation{Max Planck Institute for Plasma Physics Garching 85748 Germany}
\author{O.~Sauter}
\affiliation{Ecole Polytechnique F\'ed\'erale de Lausanne (EPFL) Swiss Plasma Center (SPC) CH-1015 Lausanne Switzerland}
\author{F.~Jenko} 
\affiliation{Max Planck Institute for Plasma Physics Garching 85748 Germany}

\begin{abstract}

Flux-tube (local) gyrokinetic codes are widely used to simulate drift-wave turbulence in magnetic confinement devices. While a large number of studies show that flux-tube codes provide an excellent approximation for turbulent transport in medium-large devices, it still needs to be determined whether they are sufficient for modeling supra-thermal particle effects on core turbulence. This is called into question given the large temperature of energetic particles (EPs), which makes them hardly confined on a single flux-surface, but also due to the radially broad mode structure of energetic-particle-driven modes. The primary focus of this manuscript is to assess the range of validity of flux-tube codes in modeling fast ion effects by comparing radially global turbulence simulations with flux-tube results at different radial locations for realistic JET parameters using the gyrokinetic code GENE. To extend our study to a broad range of different plasma scenarios, this comparison is made for four different plasma regimes, which differ only by the profile of the ratio between the plasma kinetic and magnetic pressure. The latter is artificially rescaled to address the (i) electrostatic limit and regimes with (ii) marginally stable, (iii) weakly unstable and (iv) strongly unstable fast ion modes. These energetic-particle-driven modes is identified as an AITG/KBAE via linear ORB5 and LIGKA simulations. It is found that the local flux-tube simulations can recover well the global results only in the electrostatic and marginally stable cases. When the AITG/KBAE becomes linearly unstable, the local approximation fails to correctly model the radially broad fast ion mode structure and the consequent global zonal patterns. According to this study, global turbulence simulations are likely required in regimes with linearly unstable AITG/KBAEs. In conditions with different fast ion-driven modes, these results might change.

\end{abstract}


\maketitle

%%%%%%%%%%%%%%%%%%%%%%%%%%%%%%%%%%%%%%%%%%%%%%%%%%%%%%%%%%%%%%%%%%%%%%%

\section{Introduction}

Plasma micro-instabilities are known to increase particle and energy fluxes (typically) from the plasma core towards the edge. The most common plasma micro-instability in the plasma core of present-day devices is driven by radial gradients in the ion temperature profiles, which provide free energy to these instabilities to grow and interact at different spatio-temporal scales in their nonlinear phase. This is called ion-temperature gradient (ITG) instability \cite{FRomanelli_PFB1989}. The turbulent transport driven by ITGs also indirectly affects the heating on thermal ions and electrons (e.g., efficiency of external systems, energy exchange between ions and electrons and radiative power). This is because any direct ion heating also increases the source of free energy available for ITG turbulence, strongly limiting the plasma performance. This is a major obstacle to overcome for making fusion a viable energy source where maximizing energy production and minimizing any energy input to enter into burning-plasma conditions is essential. 

An increasing number of studies show that supra-thermal particles - generated by external heating systems such as neutral beam injection (NBI) and ion cyclotron resonant frequency (ICRH) - might strongly stabilize ITG turbulence and improve the heating efficiency in experiments \cite{MRomanelli_PPCF2010,Holland_NF2012,Citrin_PRL_2013,Garcia_NF2015,DiSiena_NF_2018,Wilkie_NF_2018,DiSiena_NF_2019}. Signatures of this beneficial effect on turbulence are seen in a large number of experimental discharges \cite{Tardini_NF2007,Mantica_PRL2009,Mantica_PRL2011,Bock_NF_2017,DiSiena_PRL_2021,Han_Nature_2022,Mazzi_Nature_2022,Garcia_PPCF_2022,Citrin_PPCF_2023}. However, understanding the specific physical mechanisms behind the turbulence regulation by fast particles in experiments is rather complex, involving mutually interacting phenomena acting at different spatio-temporal scales. 

Particularly relevant insights in this regard have been obtained in the past decade by modeling experimental discharges with large fast ion content by using flux-tube (local) gyrokinetic codes, able to disentangle the complex fast ion effects on turbulence more easily. A first direct effect of the inclusion of an additional ion species is the dilution of thermal ions to ensure plasma quasi-neutrality \cite{Tardini_NF2007}. Due to the large energetic particle pressure and pressure gradients, modifications on the plasma geometry due to Shafranov-shift effects have also been shown to have a beneficial effect \cite{Bourdelle_NF_2005}. Recently, two new stabilizing mechanisms have been identified. In particular, fast ions are found to interact with the background microturbulence through a wave-particle resonance effect, when the energetic particle magnetic drift frequency gets close to the linear ITG frequencies \cite{DiSiena_NF_2018,DiSiena_PoP_2019}. This resonant stabilization has been found to be an effective way of improving the plasma confinement for ICRH schemes at ASDEX Upgrade \cite{DiSiena_PRL_2021}, JET \cite{Bonanomi_NF_2018} and, possibly, in optimized stellarators \cite{DiSiena_PRL_2020}. The other fast ion effect on turbulence recently identified is the nonlinear electromagnetic fast particle effect on ITG turbulence, first observed in Ref.~\cite{DiSiena_NF_2019}, where it has been shown that fast ions can provide linearly marginally stable MHD-type of modes, which are nonlinearly excited by an energy coupling with ITG scales and frequencies. This energy redistribution depletes the ITG drive reducing the overall transport levels. Additionally, if the energy injected into these MHD-type of modes is sufficiently large, an increase in the zonal flow amplitude is usually observed, further reducing transport levels \cite{DiSiena_JPP_2021}. 

Although the previously mentioned studies have provided major insights into the direct effect that supra-thermal particles might have on turbulent transport, they are based on flux-tube simulations. This can be a major limitation in studying energetic particle physics, where radially global effects are known to play an important role \cite{Lauber_PR_2013,Zonca_RMF_2016}. 

Flux-tube gyrokinetic codes assume that turbulent transport is essentially a local phenomenon and can be correctly described at a specific plasma location by solving the nonlinear Vlasov-Maxwell equations in a narrow radial domain. In this radial domain, constant temperatures, densities and geometrical quantities are assumed, however, keeping finite gradients (the drive of plasma instabilities) in the Vlasov equations. In addition, periodic boundary conditions are applied along the radial directions allowing a Fourier decomposition. Under these numerical simplifications, flux-tube gyrokinetic codes have a great advantage with respect to global codes since they can run with significantly lower computational resources. At the same time, flux-tube simulations are considerably more straightforward to run, and their output easier to analyze than global simulations.

Global gyrokinetic codes simulate a broad plasma radial domain, hence retaining e.g., variations of the equilibrium quantities, but not allowing for Fourier decomposition along the radial direction. Periodic boundary conditions can also not be applied in global codes and plasma profiles are sustained either via artificial Krook sources (gradient-driven) keeping profiles (on-average) fixed to the initial ones or via physical sources in flux-driven global simulations. In this manuscript all the global simulations will be performed running GENE in gradient-driven mode. A comprehensive description of the flux-tube and global versions of gyrokinetic codes can be found in Ref.~\cite{Goerler_phd_thesis}. The codes used for the linear simulations are GENE \cite{Jenko_PoP2000,Goerler_JCP2011}, ORB5 \cite{Lanti_CPC_2020, Mishchenko_CPC_2019} and LIGKA ~\cite{Lauber_JCP_2007}, while the turbulence simulations only GENE was used.

Global codes are able to capture non-local turbulent effects, such as profile shearing \cite{Garbet_PoP_1996,Waltz_PoP_2005}, turbulent avalanches \cite{Candy_PRL_2003,Sarazin_PoP_2000}, internal transport barriers \cite{Strugarek_PPCF_2013,DiSiena_PRL_2021,Di_Siena_PPCF_2022} and turbulence spreading \cite{Hahm_PPCF_2004}. However, the more complete physical description of global codes comes at the price of significantly more demanding numerical simulations in terms of computational resources, potentially less accurate numerical schemes (e.g., when treating radial derivatives) and more challenging simulations in terms of numerical resolution and setups.

Flux-tube codes have been widely used in the plasma physics community for studying drift-wave turbulence and have been shown to reproduce the radially global results in the limit of $1/\rho^* \rightarrow \infty$ for adiabatic electron simulations \cite{McMillan_PRL_2010}. Here, $\rho^* = \rho/a$ represents the ratio between the ion Larmor radius ($\rho$) and the minor radius ($a$) of the device. Naturally, the validity of flux-tube gyrokinetic codes is called into question in plasma regimes where non-local effects might play a significant role in setting the correct turbulent levels. This is potentially the case for energetic particles and their interaction with drift-wave turbulence. This is because energetic particle driven modes have a particularly broad mode radial structure that can be captured only by retaining the full radial dependencies of the equilibrium quantities. On the other hand, due to the large energetic particle temperatures, their Larmor radius is particularly large, potentially pushing the system far from the condition $1/\rho_f^* \rightarrow \infty$ (with $\rho_f$ fast ion Larmor radius) where flux-tube and global simulations are expected to provide similar results.

Therefore, it is important to assess to which extent flux-tube simulations represent an adequate approximation of the more challenging global runs,  given the different levels of complexity and the reduced numerical cost of running a single flux-tube simulation with respect to a radially global one. This is essential for better modeling and understanding the underlying fast ion physics and identifying the range of parameters which maximize their beneficial effect on turbulence. This is of high relevance in particular for future fusion reactors, not only in order to achieve a more reliable prediction of their performance, but also - if corresponding actuators could be devised - to develop a possible way to improve it.

This is addressed in the present manuscript for realistic plasma parameters inspired by the L-mode JET discharge $\# 73224$ \cite{Mantica_PRL2011,Citrin_PRL_2013} for energetic particles generated via heating schemes. We found that the flux-tube simulations recover the global turbulent fluxes qualitatively well only when energetic particle driven modes are marginally stable. These modes are identified as a mixture of Alfv\'enic ion temperature gradient modes (AITGs) \cite{Zonca_1998,Hayward-Schneider_NF_2022} and kinetic beta-induced Alfv\'en eigenmodes (KBAEs) \cite{Heidbrink_PoP_1999}. When these modes are linearly destabilized, the flux-tube simulations predict a substantial increase in the electron and fast ion turbulent fluxes (and with a smaller degree of the thermal ion fluxes) not in agreement with the global findings. This is explained when comparing the zonal radial electric field and zonal currents, showing that zonal structures arising due to the interaction with energetic particle modes are not captured correctly by flux-tube codes. Another important observation is that the global simulation with marginally stable energetic particle modes shows similar features of the flux-tube ones, suggesting that the physical interpretation summarized above - primarily based on flux-tube simulations - involving marginally stable energetic particle modes suppressing ion-scale turbulent transport \cite{DiSiena_NF_2019,DiSiena_JPP_2021} might still hold in radially global setups. At higher fast ion energies, the validity of the local approximation might further decrease since (i) the smaller value of $1/\rho^*_{EP}$ causes the description of fast ion transport to become more global \cite{McMillan_PRL_2010}, and secondly, various energetic particle driven modes with a wider mode structure may be triggered \cite{Mishchenko_PoP_2009,DiSiena_JPP_2021}, resulting in less accurate linear description of drive/damping in flux-tube models. This will be addressed in future works.

This paper is organized as follows. The JET plasma scenario inspiring the numerical setup used in this manuscript is discussed in Section \ref{sec1}. The numerical setup and the different plasma regimes investigated with flux-tube and global simulations are summarized in Section \ref{sec2}. While Section \ref{sec3} presents a linear comparison between flux-tube and global GENE simulations, Section \ref{sec4} shows a linear benchmark between global GENE and ORB5 simulations for a simplified scenario (with minor differences on the magnetic equilibrium, reduced ion-to-electron mass ratio and neglecting collisions). Here, the nature of the energetic particle driven mode is identified. The global nonlinear GENE simulations are discussed in Section \ref{sec5}. The nonlinear comparison between flux-tube and global GENE simulations is performed in Section \ref{sec6} - \ref{sec8} for the setup inspired by the JET discharge. We compare the turbulent fluxes (Section \ref{sec6}), frequency spectra of the electrostatic potential (Section \ref{sec7}) and zonal radial electric field and zonal currents (Section \ref{sec8}). Finally, conclusions are drawn in Section \ref{sec9}.

\section{JET L-mode scenario} \label{sec1}

\begin{figure*}
\begin{center}
\includegraphics[scale=0.35]{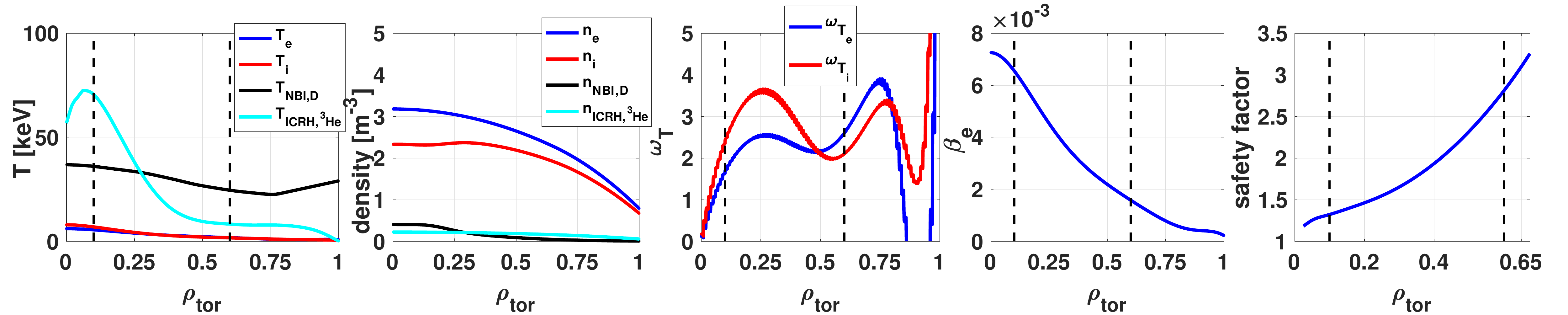}
\par\end{center}
\caption{Radial profile of a) temperatures and b) densities of each plasma species, c) thermal ion and electron logarithmic temperature gradient $\omega_{T} = -a d$ln$(T)/d \rho_{tor}$ with $a$ minor radius of the device, d) ratio between the electron kinetic and ratio between the electron kinetic and magnetic pressure $\beta_e = 8\pi n_e T_e / B_0^2$ with $B_0$ the on-axis toroidal magnetic field and e) safety factor.}
\label{fig:fig1}
\end{figure*}

The reference scenario selected for these analyses is inspired by the L-mode discharge $\# 73224$ (with Carbon wall) \cite{Mantica_PRL2011} (the differences between the setup employed in this manuscript and this JET discharge are explained in Section \ref{sec2}). This is a deuterium plasma externally heated via 11 MW of neutral-beam (NBI) and 3 MW of ion cyclotron resonance heating (ICRH) power with $^3$He minority. The on-axis magnetic field is $B_0 = 3.36$T, the plasma current $I_p = 1.8$MA with the minority concentration kept at $\sim 6\% - 8\%$ respect to the electron density. While the thermal profiles are reconstructed via CRONOS \cite{Artaud_NF_2010} simulations, the energetic particle ones are calculated by NEMO/SPOT \cite{Schneider_NF_2011} and SELFO \cite{Hedin_NF_2002}, respectively, for the NBI and ICRH generated fast particles. The resulting temperature and density profiles are illustrated in Fig.~\ref{fig:fig1} together with the logarithmic thermal ion temperature gradient ($\omega_{T_i} = -(1/T_i)dT_i/d\rho_{tor}$), the ratio between the electron kinetic and magnetic pressure ($\beta_e$) and the safety factor. When looking at Fig.~\ref{fig:fig1}, we can clearly observe a sharp increase of $\omega_{T_i}$ in the radial domain $\rho_{tor} = [0.2 - 0.4]$. Detailed flux-tube gyrokinetic simulations were performed on this specific scenario at the radial position $\rho_{tor} = 0.33$, linking the improvement of plasma confinement to a turbulence suppression via nonlinear electromagnetic fast particle effects. 
In this manuscript, the previous studies employing flux-tube simulations are extended by means of radially global turbulence simulations with the main focus on understanding the limits of local codes in modeling supra-thermal particle effects on plasma turbulence.

\section{Simulation details} \label{sec2}

The numerical GENE simulations are performed by retaining thermal deuterium, electrons (with realistic ion-to-electron mass ratio) and the NBI-generated fast deuterium. Carbon impurities and ICRH-generated fast particles ($^3$He) are neglected to reduce the otherwise prohibitive computational resources required, despite their presence being essential for matching power balance in GENE local simulations \cite{Citrin_PRL_2013}. This is not a major limitation of the present study, which aims to assess the range of validity of flux-tube (local) gyrokinetic codes in modeling energetic particle effects on core turbulence correctly and not in reproducing the experimental findings. These analyses are done by comparing the turbulent fluxes radial profile of the GENE global simulations with the ones obtained with the flux-tube (local) code at different radial locations for each species. 
\begin{figure}
\begin{center}
\includegraphics[scale=0.4]{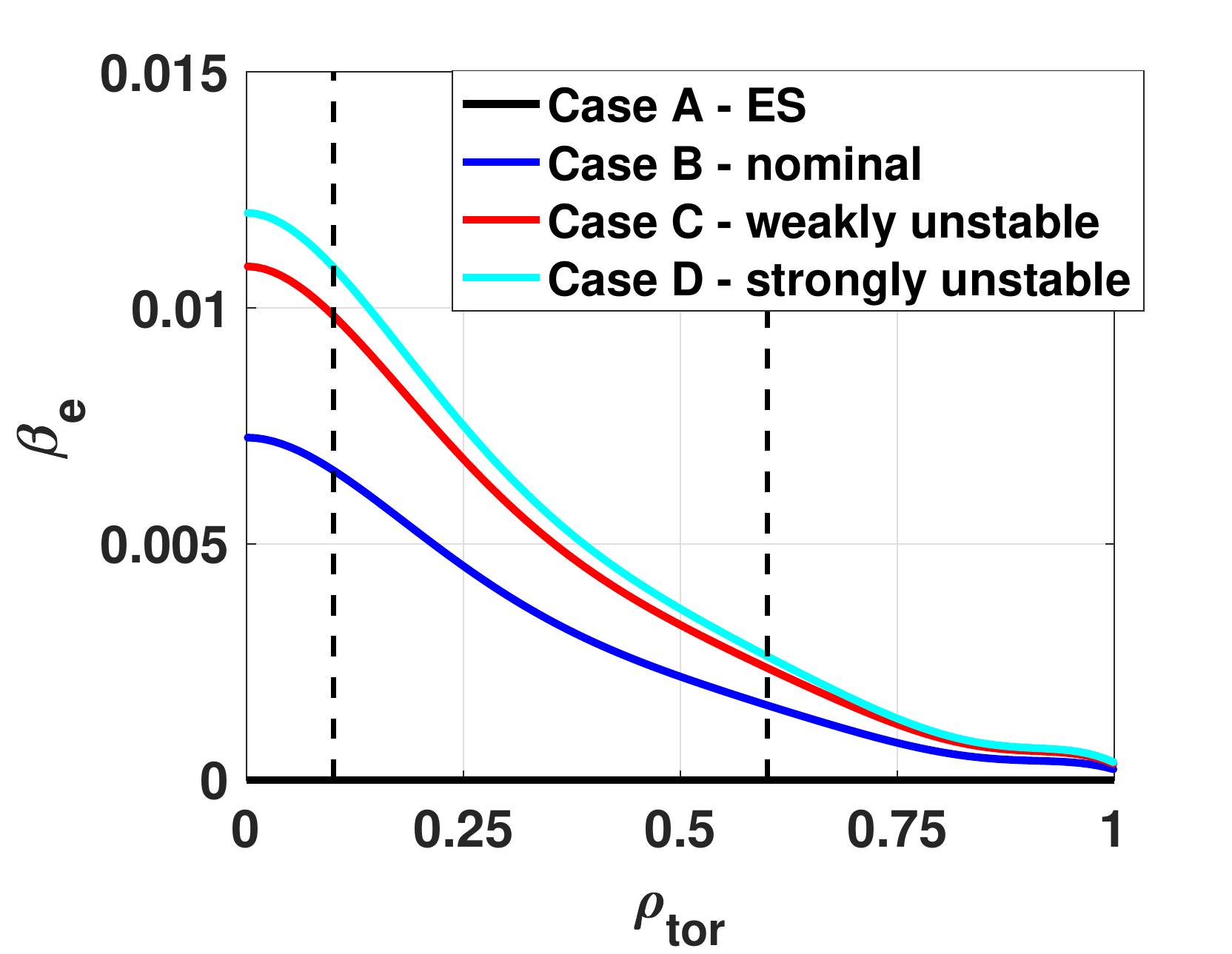}
\par\end{center}
\caption{Ratio between the electron kinetic and magnetic pressure ($\beta_e$) profile characterizing the different regimes considered for in this manuscript. These profiles are obtained from the nominal scenario (Case B) by re-scaling $\beta_e$ by a constant factor. The thermal and fast ion $\beta$ profiles are modified accordingly by the same factor. Here, the abbreviation "ES" stands for electrostatic, while EPM stands for energetic particle mode (its nature is identified in Sec.~\ref{sec5}).}
\label{fig:fig2}
\end{figure}

To cover a broad range of plasma conditions, this comparison is performed by selecting four different setups that differ only by the plasma beta profile employed in the GENE simulations. These modifications were made by artificially re-scaling the total beta profile in the Vlasov and field equations of GENE without changing the pressure profiles. The magnetic equilibrium is kept fixed to the nominal one regardless of the beta profile employed. Fig.~\ref{fig:fig2} shows the electron plasma beta ($\beta_e$) associated with these different scenarios. In particular, we considered (i) $\beta_e = 1e-4$ at $\rho_{tor} = 0.35$ (electrostatic limit - labelled Case A in Fig.~\ref{fig:fig2}), (ii) $\beta_e = 0.33e-2$ at $\rho_{tor} = 0.35$ (nominal $\beta_e$ - labelled Case B in Fig.~\ref{fig:fig2}), (iii) $\beta_e = 0.48e-2$ at $\rho_{tor} = 0.35$ (weakly unstable AITG/KBAEs, as discussed in Sec.~\ref{sec3} - labelled Case C in Fig.~\ref{fig:fig2}) and (iv) $\beta_e = 0.53e-2$ at $\rho_{tor} = 0.35$ (strongly unstable AITG/KBAEs, as discussed in Sec.~\ref{sec3} - labelled Case D in Fig.~\ref{fig:fig2}). The location $\rho_{tor} = 0.35$ is the centre of the radial domain simulated by GENE in the global simulations.

%\subsection{Code description} \label{sec2a}

\subsection{Numerical setup and resolution} \label{sec2b}

The following section summarises the numerical parameters and grid resolutions employed for the radially global and flux-tube (local) GENE simulations. Realistic ion-to-electron mass ratio, electromagnetic effects (with the only exception of Case A where $\beta_e = 1e-4$), collisions modeled with a linearized Landau operator with energy and momentum conserving terms and realistic magnetic equilibrium reconstructed via TRACER-EFIT \cite{Xanthopoulos_PoP_2009} are considered. The radial domain retained in the GENE global simulations is $\rho_{tor} = [0.1-0.6]$, thus covering the fraction of the plasma volume where the logarithmic ion temperature profile increases the most in the experimental discharge. As extensively discussed in Ref.~\cite{Mantica_PRL2011,Citrin_PRL_2013}, this is also the radial region where electromagnetic fast particle effects on turbulent transport are more significant according to flux-tube simulations.

\begin{figure}
\begin{center}
\includegraphics[scale=0.45]{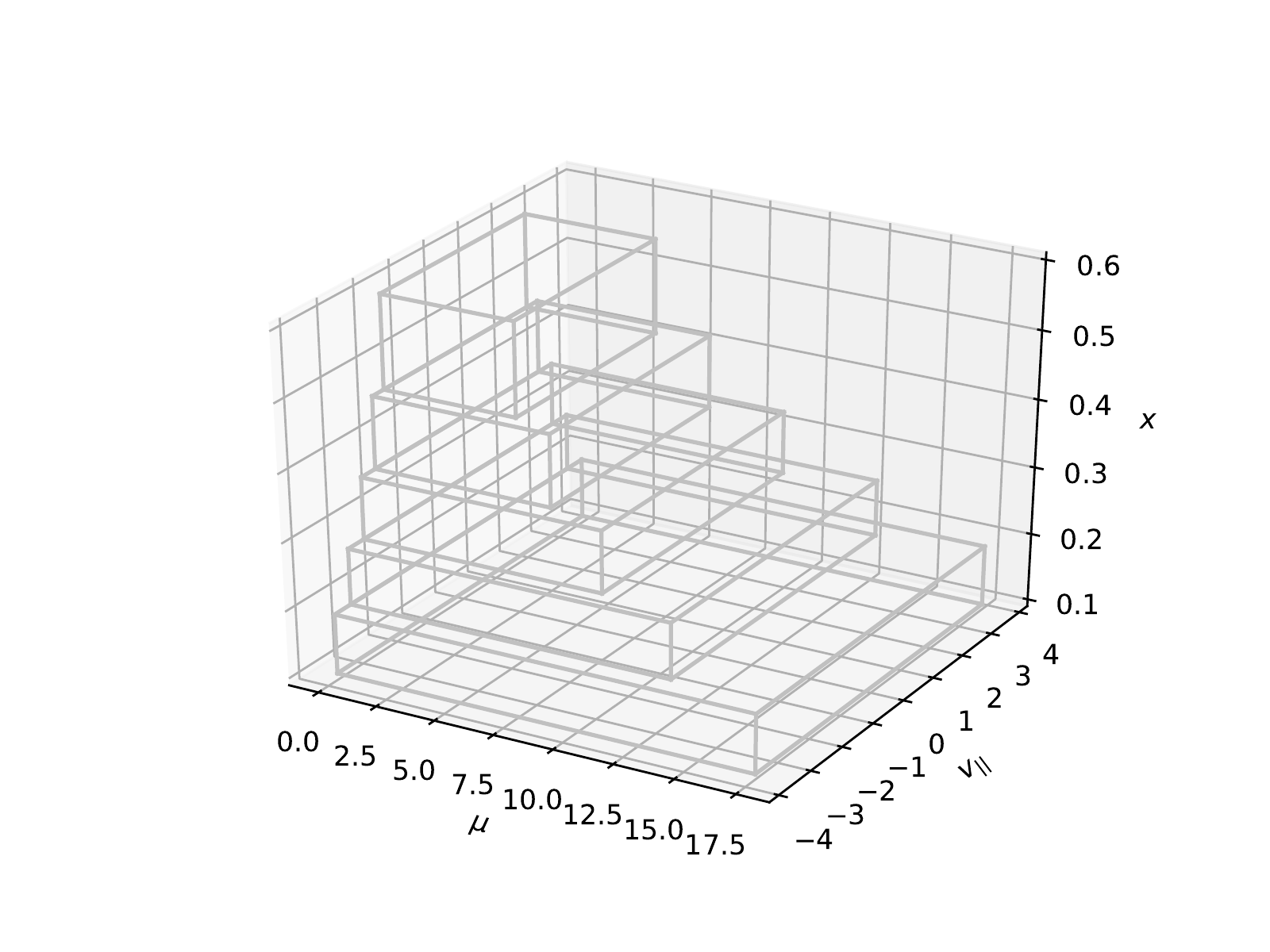}
\includegraphics[scale=0.4]{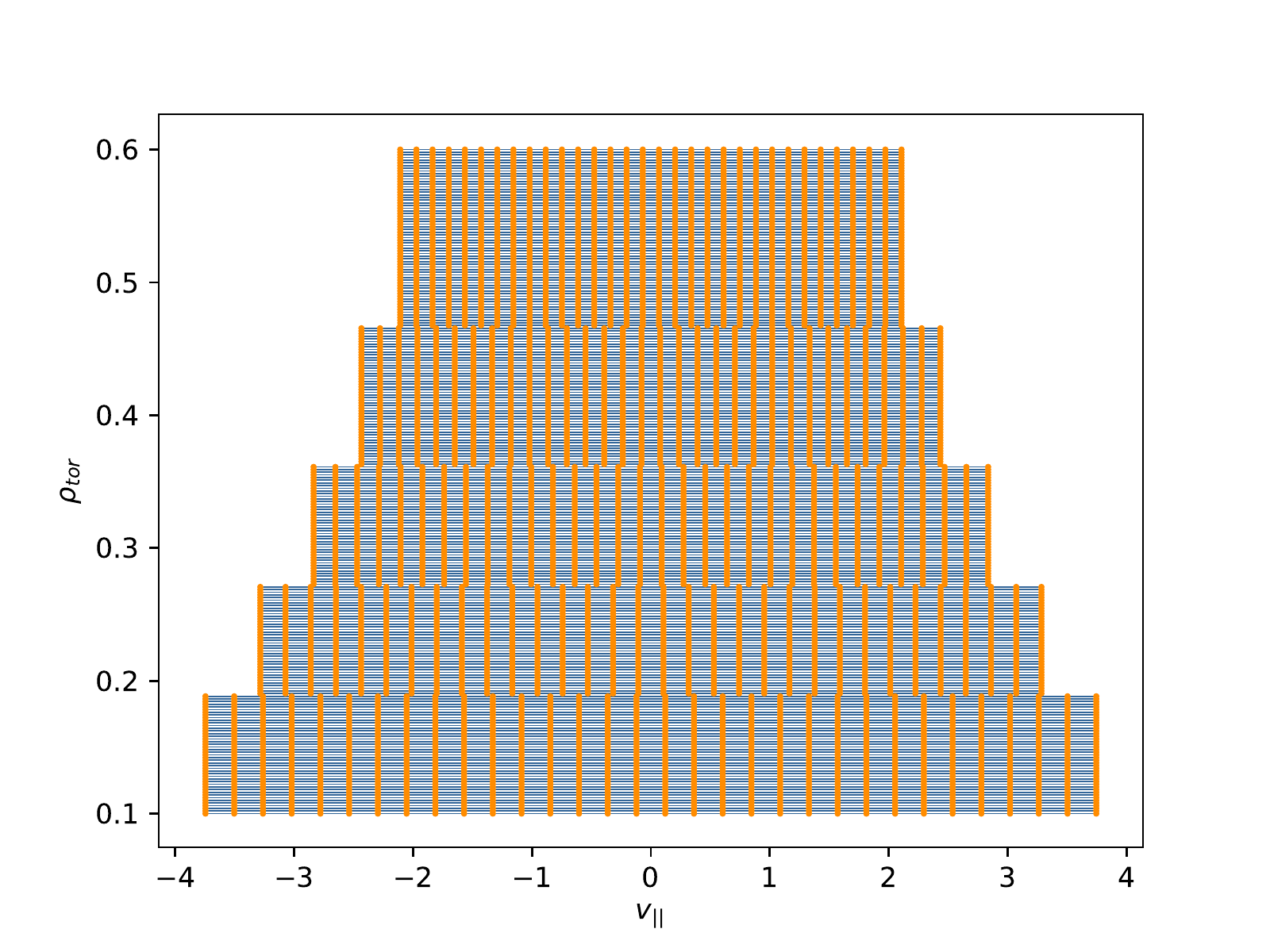}
\includegraphics[scale=0.4]{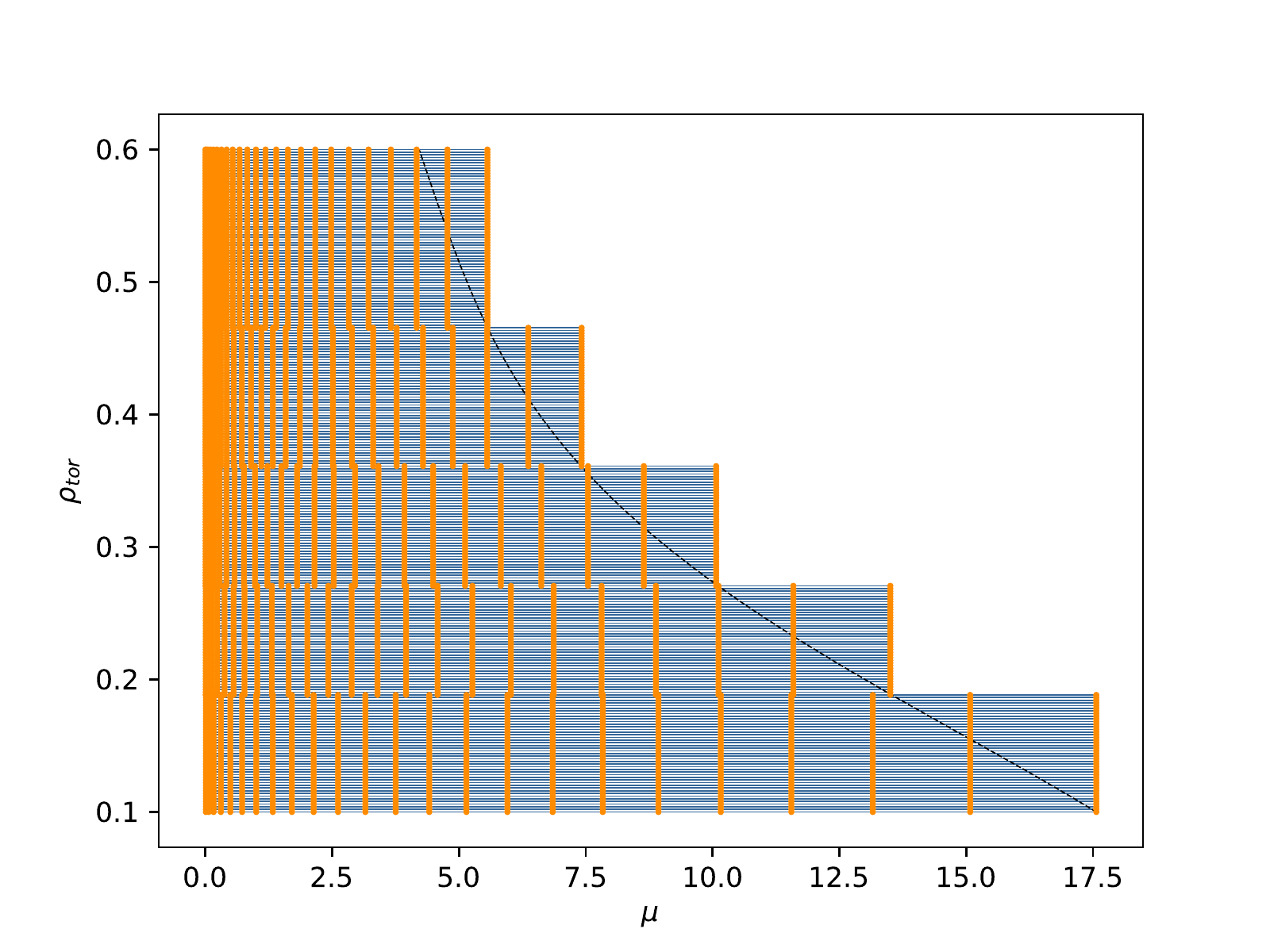}
\par\end{center}
\caption{Radially dependent block-structured velocity grid with five blocks in a) $(\rho_{tor}, v_\shortparallel,\mu)$ and its slices in b) $(\rho_{tor}, v_\shortparallel)$ and c) $(\rho_{tor},\mu)$. The vertical orange lines in each blocks of b) and c) represent the $(v_\shortparallel,\mu)$ grid points used in each directions.}
\label{fig:fig3}
\end{figure}

All the radially global simulations presented within this manuscript are performed running GENE in the gradient-driven mode. Therefore, we applied Krook particle and heat sources/sinks to keep the plasma profiles (on-average) fixed to the initial ones. The amplitude of the Krook particle ($\gamma_p$) and heat ($\gamma_k$) coefficients used are, respectively, $\gamma_p = 1$ and $\gamma_k = 0.05$ in units of $c_s / a$, where $c_s = (T_e / m_i)^{1/2}$ is the sound speed, with $T_e$ the electron temperature at the centre of the radial domain, $a$ the minor radius and $m_i$ the bulk ion mass in proton units. These have been selected after performing nonlinear scans over the Krook amplitude and retaining the minimum values keeping the plasma profiles close to the initial ones. This minimizes nonphysical effects due to the large artificial sources \cite{Mariani_PPCF_2019}. Dirichlet boundary conditions are enforced in the global version of the code GENE to damp fluctuations in the so-called buffer regions, covering $10\%$ of the simulated radial domain. Within these regions, a Krook-type operator is applied with a relaxation rate of $\gamma_k = 1.0 c_s/a$. Fine electron scale turbulence due to electron temperature gradient (ETG) modes is artificially damped in the global GENE simulations via numerical fourth order hyperdiffusion. To allow for a more direct comparison between the flux-tube and global results, fluctuations of the magnetic field along the field-aligned direction are neglected in the underlying GENE equations since these terms are not yet implemented in the global code. Furthermore, a radially dependent block-structured grid is employed in the global simulations to reduce the velocity resolution required to resolve correctly the dynamics of all the plasma species \cite{Jarema_CPC_2016}. This block-structured grid is shown in Fig.~\ref{fig:fig3}a in the $(v_\shortparallel,\mu,\rho_{tor})$-space, in Fig.~\ref{fig:fig3}b for its slice at $\mu=0$ and in Fig.~\ref{fig:fig3}c at the plane at $v_\shortparallel = 0$.

The grid resolutions and box sizes employed for the global and flux-tube (local) simulations are summarized in table \ref{table:tab1}. The discretized toroidal mode number is given by $n = n_{0,min} \cdot j$ with $j$ being integer-valued in the range $j = [0,1,2, ..., n_{ky0}]$. 
\begin{table}
\centering
\begin{tabular}{ ||c| c| c|| }
 \hline
 GENE setup & flux-tube (local) & global\\
 \hline
 $n_x \times n_{ky0} \times n_z$   & $250 \times 64 \times 24$    & $(500 - 250) \times 64 \times 32$\\
 $n_{v_\shortparallel} \times n_{\mu}$   & $32 \times 16$    &$48 \times 24$ \\
  $n_{0,min}$    &  2  &2   \\
  $l_{v_{min}}, l_{v_{max}}$    &  -3,  3  &  -3.75, 3.75\\
  $l_{w_{min}}, l_{w_{max}}$    &  0, 9 &  0, 17.6 \\
 \hline
\end{tabular}
\caption{GENE grid resolution and box sizes. Here, $n_x$, $n_{ky0}$ and $n_z$ denote, respectively, the numerical resolution used along the radial, bi-normal and field-aligned directions; $n_{v_\shortparallel}$ and $n_{\mu}$ the resolution used in velocity space for the velocity component parallel to the background magnetic field and the magnetic moment; $l_{v_{min}}, l_{v_{max}}$ and $l_{w_{min}}, l_{w_{max}}$ denote, respectively, the extension (minimum/maximum value) of the simulation box in the $v_\shortparallel$ and $\mu$ directions in units of the thermal velocity of each species. While the thermal velocity is computed at the considered location for flux-tube runs, it is taken at the centre of the radial grid in the global simulations.}
\label{table:tab1}
\end{table}
Dedicated convergence studies have been performed for both flux-tube and global simulations.

\section{Linear flux-tube and global comparison} \label{sec3}

\begin{figure}
\begin{center}
\includegraphics[scale=0.16]{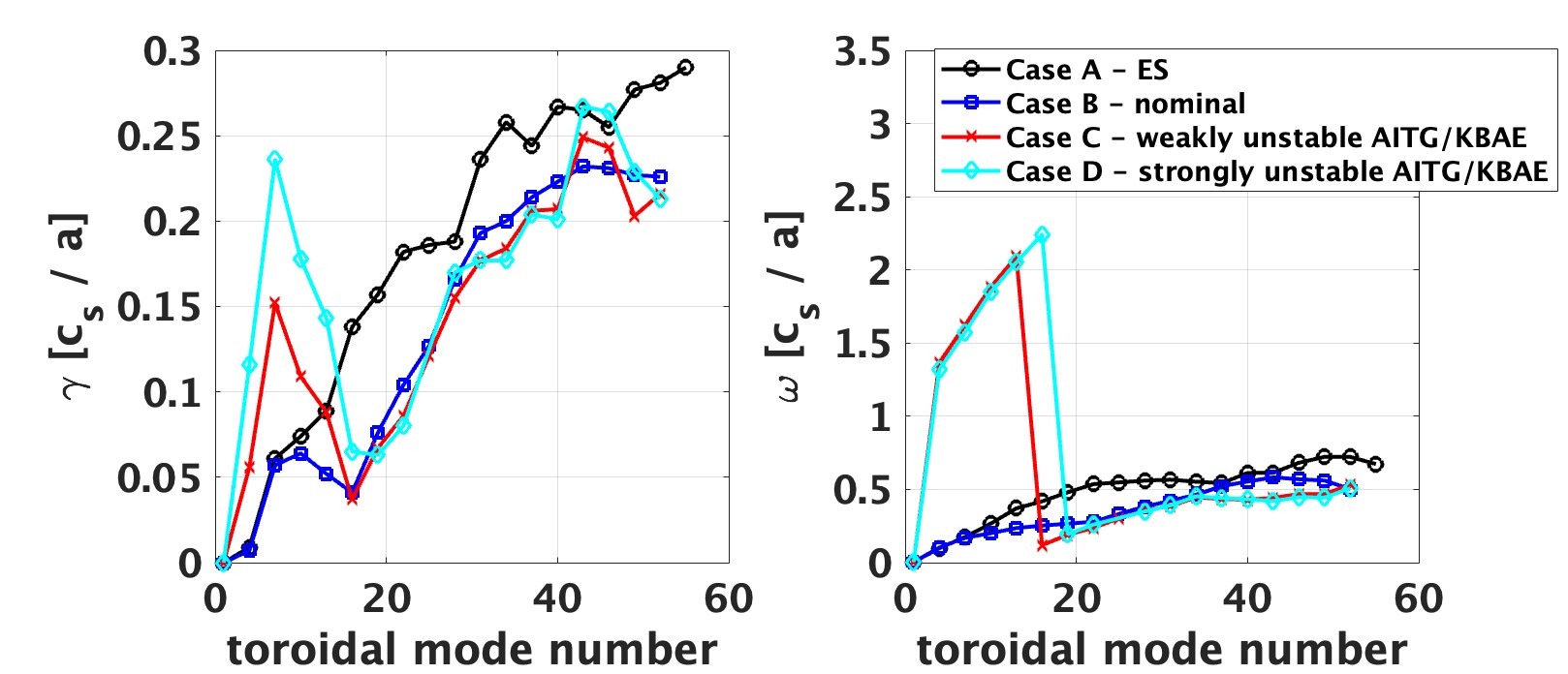}
\par\end{center}
\caption{Linear growth rates a) and frequencies b) of the most unstable modes at different toroidal mode numbers for the different plasma scenarios described in Fig.~\ref{fig:fig2} from global GENE simulations.}
\label{fig:fig4}
\end{figure}

\begin{figure}
\begin{center}
\includegraphics[scale=0.30]{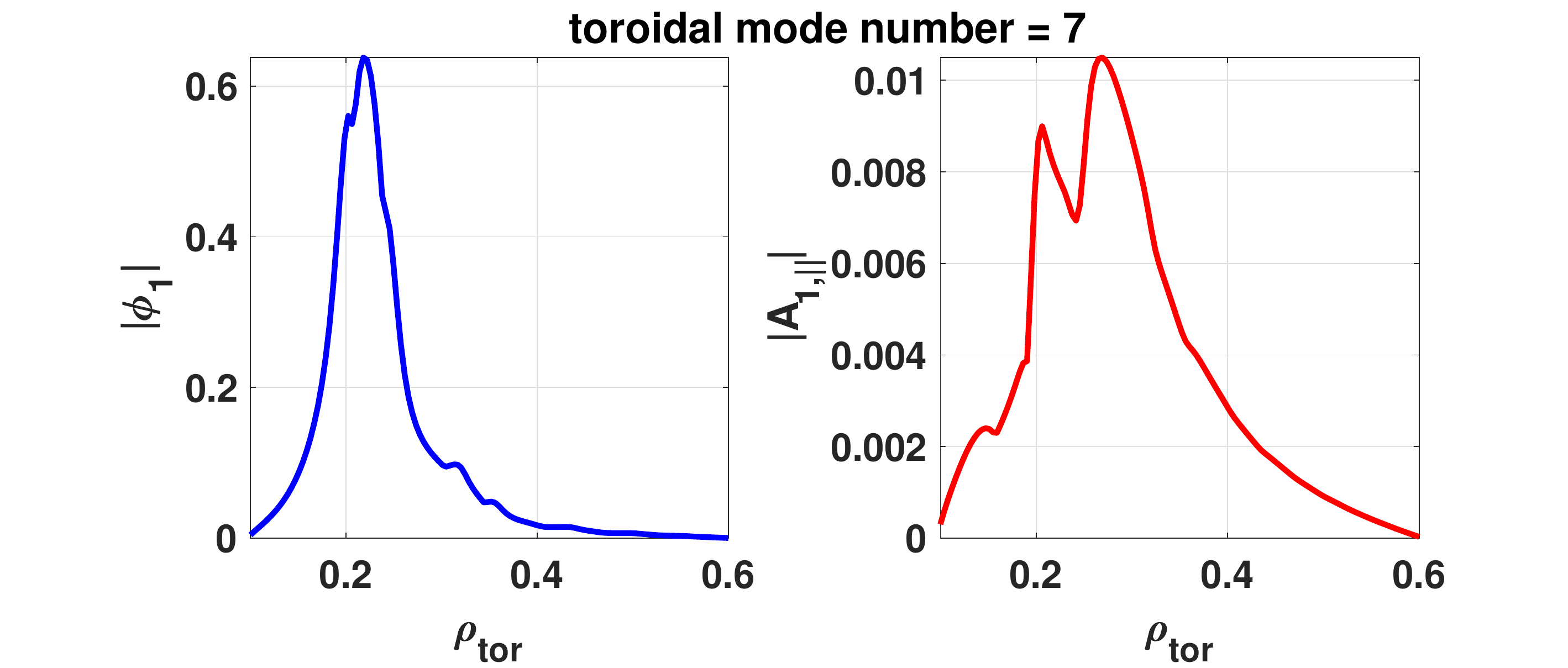}
\par\end{center}
\caption{Radial structure of the electrostatic (left) and magnetic (right) potentials averaged over the field-line coordinate obtained in the global GENE simulations for Case D at the toroidal mode number $n = 7$, corresponding to the most unstable AITG/KBAE.}
\label{fig:fig5}
\end{figure}

\begin{figure}
\begin{center}
\includegraphics[scale=0.40]{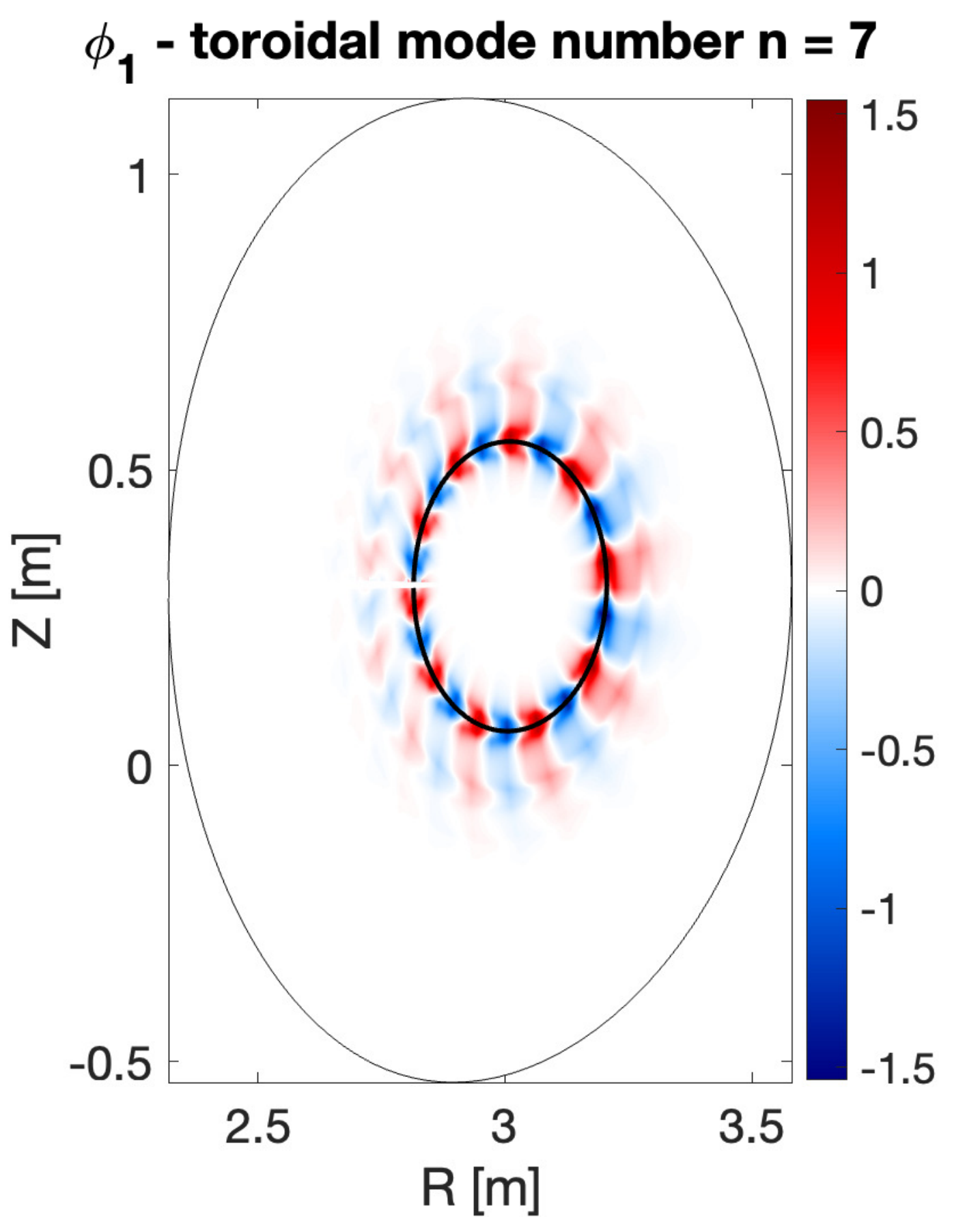}
\par\end{center}
\caption{Poloidal cross section of the electrostatic potential for the GENE global simulation at $n = 7$. The solid black lines delimit, respectively, the location of $q = 10/7$ (inner line) and the flux-surface at $\rho_{tor} = 0.6$ (outer line).}
\label{fig:fig6}
\end{figure}

We begin our analyses by performing linear stability analyses for the four different plasma scenarios introduced above. These simulations are done by retaining the effect of collisions, realistic ion-to-electron mass ratio and finite-$\beta_e$ with the global version of the code GENE. The results are summarized in Fig.~\ref{fig:fig4}, where the linear growth rates and frequencies are illustrated for different toroidal mode numbers. 
While Case A and Case B are dominated by ITG micro-instabilities only, we observe a sharp mode transition for $n < 18$ in the frequency range $\omega [c_s/a] \sim [1.5 - 2]$ for Case C and Case D. These cases have been constructed ad-hoc to linearly destabilize a fast ion-driven mode at low toroidal mode numbers. The growth rate of this mode peaks at $n = 7$ and increases with $\beta_e$. Moreover, this energetic particle mode is located at $\rho_{tor} \sim 0.23$, as shown when looking at the radial mode structure of the electrostatic and magnetic potentials, see Fig.~\ref{fig:fig5}.
When looking at the poloidal cross section of the electrostatic potential for the toroidal mode number $n = 7$ (see Fig.~\ref{fig:fig6}), we can clearly observe two dominant poloidal mode numbers, i.e., $m_1 = 10$ and $m_2 = 11$. 

\begin{figure}
\begin{center}
\includegraphics[scale=0.16]{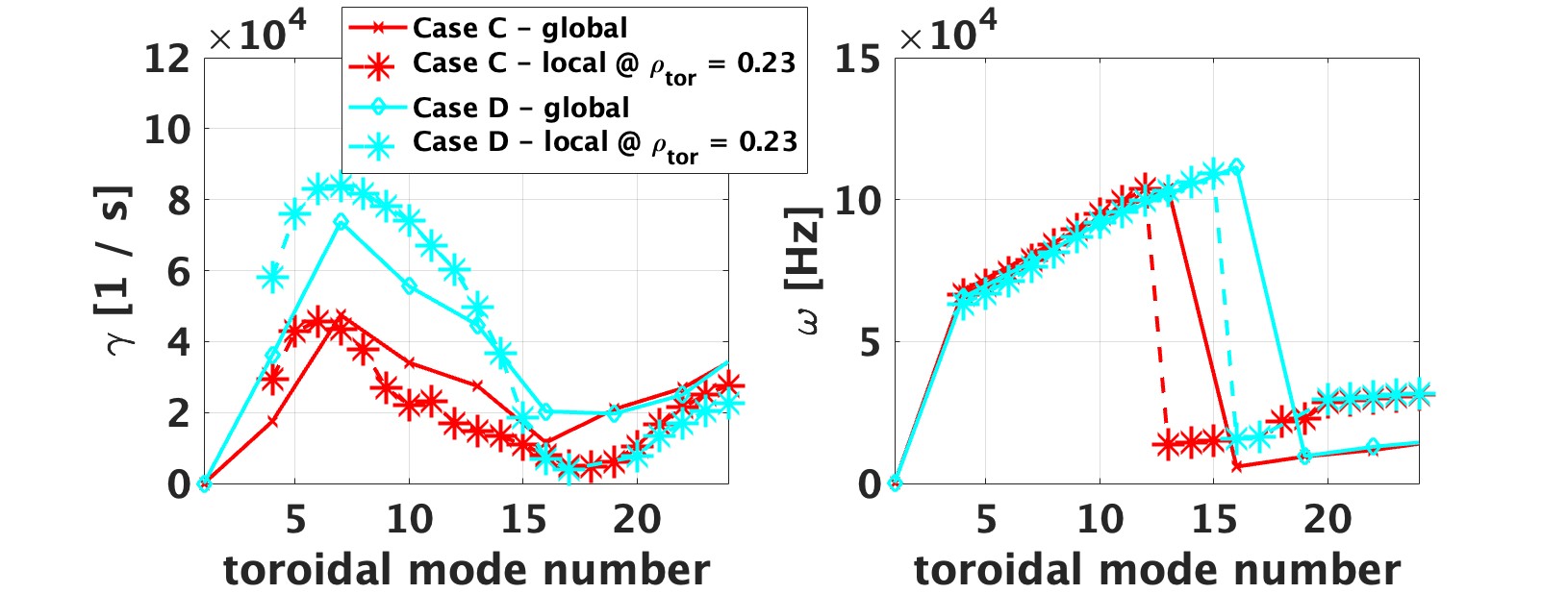}
\par\end{center}
\caption{Comparison of the linear a) growth rates and b) frequencies obtained in global (continuous line) and flux-tube (dotted line) simulations at $\rho_{tor} = 0.23$ for Case C and Case D at different toroidal mode numbers.}
\label{fig:fig7}
\end{figure}

At the radial position where this Alfv\'en mode peaks ($\rho_{tor} \sim 0.23$), flux-tube linear simulations are performed for Case C and Case D and the linear growth rates and frequencies compared with the results of Fig.~\ref{fig:fig4}. Given the different normalization between the local (reference parameters taken at $\rho_{tor} = 0.23$) and the global (reference parameters taken at the centre of the radial domain $\rho_{tor} = 0.35$) codes, the growth rates and frequencies are expressed in physical units. The results are illustrated in Fig.~\ref{fig:fig7}. 
Interestingly, we find a good qualitative agreement between the flux-tube and global results with differences of $\sim 20\%$ for the growth rates and $\sim 2\%$ for the mode frequencies. The larger differences observed when comparing the growth rates and frequencies of the ITG modes at $n > 12$ are caused by the ITG modes peaking at a different radial location and, hence, not entirely captured by the flux-tube simulations centred at $\rho_{tor} = 0.23$. The Alfv\'en mode observed in this Section is identified as a mixture of Alfv\'enic ion temperature gradient modes (AITGs) \cite{Zonca_1998,Hayward-Schneider_NF_2022} and kinetic beta-induced Alfv\'en eigenmodes (KBAEs) \cite{Heidbrink_PoP_1999} (see Section \ref{sec4} for more details).

These findings are consistent with Ref.~\cite{DiSiena_JPP_2021}, showing that GENE flux-tube recovers well the linear growth rates and frequencies of the ITPA benchmark case for a toroidal-Alfv\'en-eigenmodes (TAE) mode located at the centre of the shear shear Alfv\'en waves (SAW) gap and far from the continuum. As discussed in detail in Ref.~\cite{DiSiena_JPP_2021}, the flux-tube code fails to reproduce the linear growth rates and frequencies correctly when the TAE undergoes a mode transition to an energetic particle mode (EPM) where phase-mixing effects are not negligible and hardly captured by a local approach.

\section{Linear benchmark with ORB5 and Alfv\'en eigenmode characterization} \label{sec4}
%\gcomment{TWHS: the model is best described in MishchenkoCPC2019. In short, it's mixed-variable ($v_\shortparallel$ not $p_\shortparallel$). I will modify the description below later to make it accurate.}

In this Section we compare the linear growth rates and frequencies obtained with the global version of GENE with the gyrokinetic code ORB5. ORB5 is a radially global Lagrangian electromagnetic particle-in-cell (PIC) code, which solves the Vlasov-Maxwell system of equations using the mixed-variable/pullback approach~\cite{Mishchenko_CPC_2019} in $(\vec{R},v_\parallel,\mu)$ coordinates, and solves the fields ($\phi$, $A_\parallel$) using a Fourier-filtered finite element method. Here, $\vec{R}$ represent the gyrocenter position, $v_\parallel$ the parallel velocity, and $\mu$ the magnetic moment. For a detailed description of the code and its numerical schemes we refer the readers to Refs.~\cite{Lanti_CPC_2020, Mishchenko_CPC_2019}. The global version of GENE has already been benchmarked with ORB5 in both linear and nonlinear simulations performed on different plasma regimes, covering ITG \cite{Goerler_PoP_2016,Dominski_PoP_2017,Merlo_PoP_2018}, TEM \cite{Dominski_PoP_2017}, GAM \cite{Biancalani_PoP_2017}, EGAM \cite{DiSiena_NF_2018_egam} and KBM \cite{Goerler_PoP_2016} physics. Here, we further extend this comparison to the scenario introduced in Section \ref{sec1} for Case D with an unstable Alfv\'en eigenmode. This is the regime with the most unstable fast ion mode, as previously observed in Fig.~\ref{fig:fig4}. The plasma profiles are the ones discussed in Section \ref{sec2}.

To simplify the numerical setup, collisions are neglected in the modeling and a reduced ion-to-electron mass ratio ($m_i / m_e = 400$) is considered. These simplifications are not expected to affect the Alfv\'en eigenmode linear properties. Only for the numerical results discussed within this Section, a different magnetic equilibrium is considered. It has been reconstructed with CHEASE \cite{Lutjens_CPC_1996} by fixing the plasma pressure, safety factor profile and last-closed flux surface to the ones describing the EQDSK file used for the previous GENE simulations. This is required since ORB5 does not have an interface to EQDSK files.
\begin{figure}
\begin{center}
\includegraphics[scale=0.4]{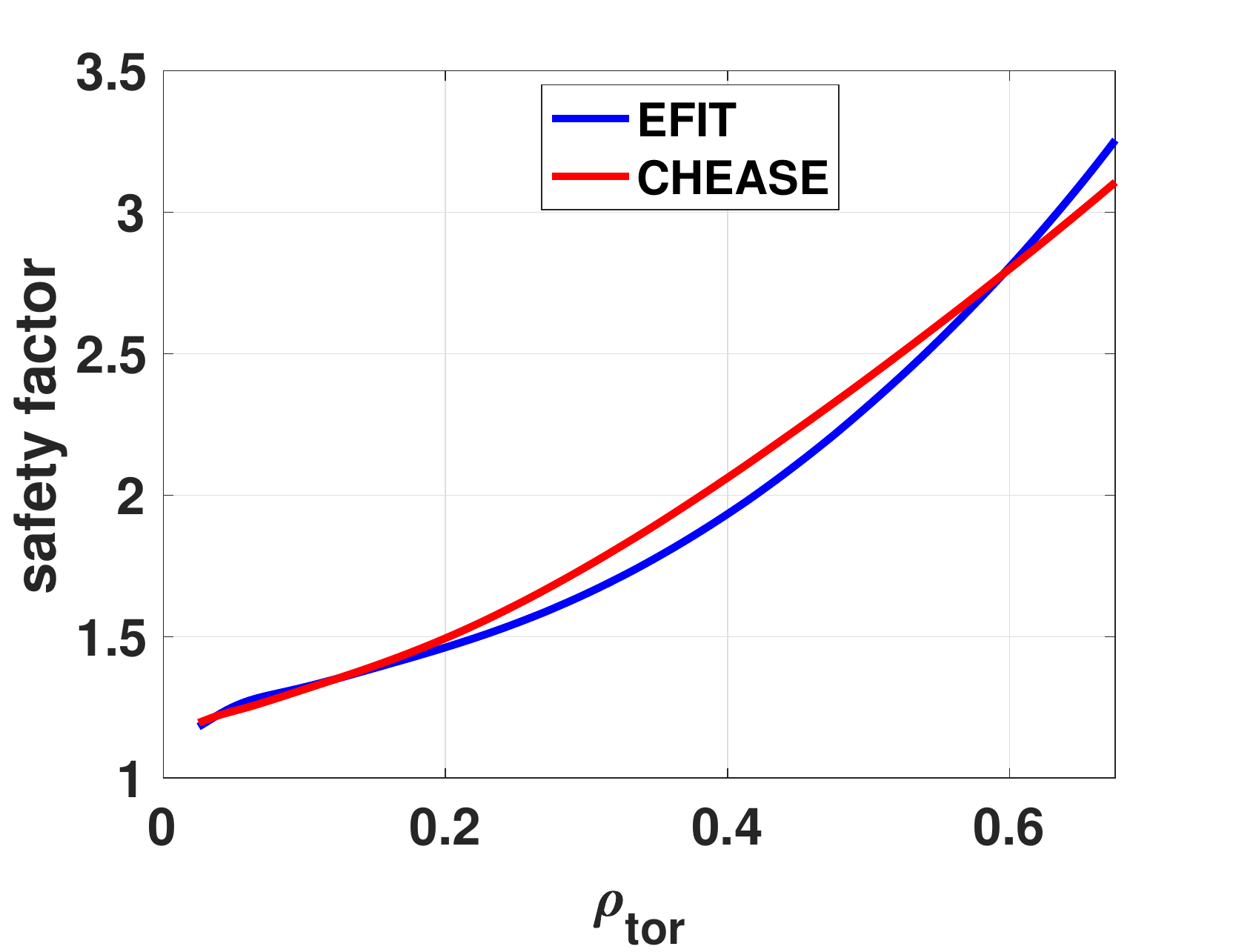}
\par\end{center}
\caption{Comparison of the safety factor profiles reconstructed via TRACER-EFIT (blue line) and extracted from the CHEASE equilibrium (red line).}
\label{fig:safety}
\end{figure}

Convergence in CHEASE is achieved only after applying a smoothing on the safety factor profile especially close to the magnetic axis (this region is anyway not considered in the GENE global simulations). For completeness we show in Fig.~\ref{fig:safety} a comparison of the safety factor profiles used only for this specific benchmark (CHEASE) and the one used for the rest of this paper (EFIT). The latter is the same as the safety factor profile shown in Fig.~\ref{fig:fig1}d. The radial domain considered in this Section is $\rho_{tor} = [0 - 0.65]$, which corresponds to a radial grid in ORB5 going from $s = [0 - 1]$.
\begin{figure}
\begin{center}
\includegraphics[scale=0.32]{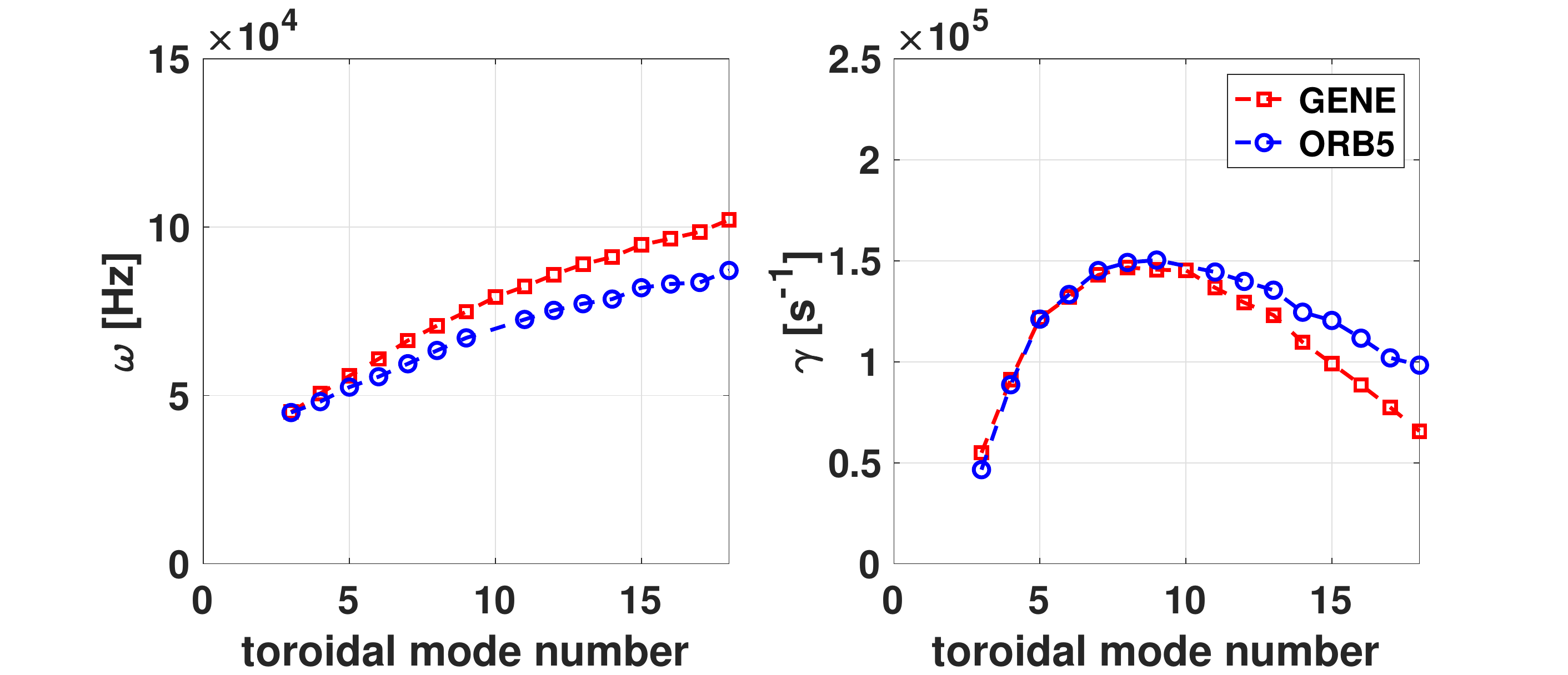}
\par\end{center}
\caption{Comparison of the linear a) growth rates and b) frequencies obtained from the radially global GENE (red squares) and ORB5 (blue circles) simulations at different toroidal mode numbers for the numerical setup characterizing Case D.}
\label{fig:lin_orb}
\end{figure}

The linear growth rates and frequencies obtained with GENE (global) and ORB5 are illustrated in Fig.~\ref{fig:lin_orb}, which shows a good quantitative agreement between the GENE and ORB5 results, with a maximum relative discrepancy of $\sim 10\%$. Notably, we observe more significant differences at higher toroidal mode numbers. This may stem from the distinct treatments of gyro-averages in the field equation employed by GENE and ORB5. Specifically, GENE adopted the full gyro-average matrices, whereas ORB5 relied on the long wavelength approximation for the simulations presented in this Section.

\begin{figure}
\begin{center}
\includegraphics[scale=0.45]{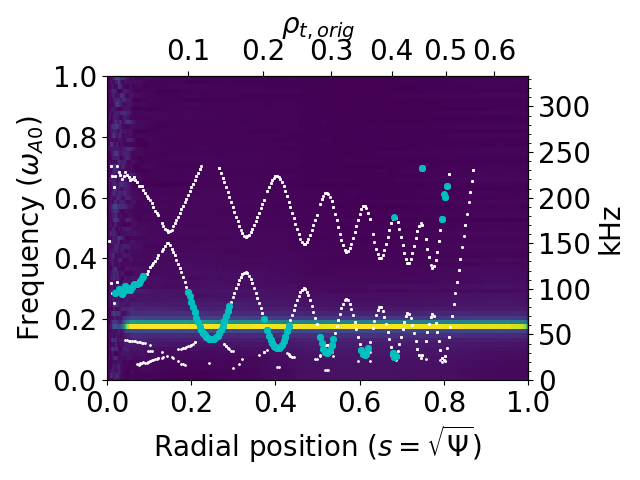}
\includegraphics[scale=0.45]{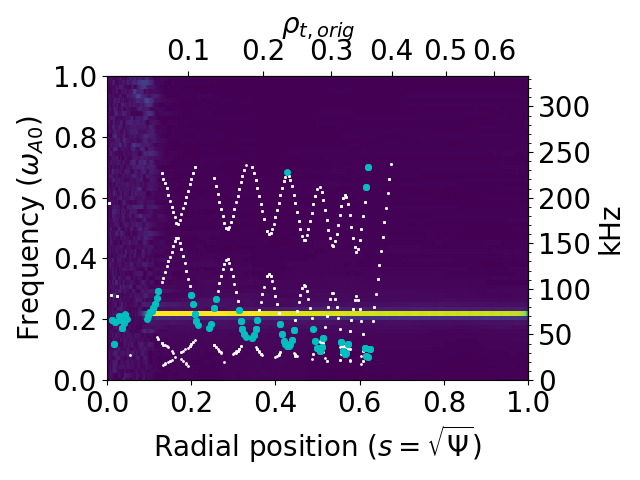}
\par\end{center}
\caption{Radial frequency spectrogram for the electrostatic potential at $n = 7$ (upper figure) and $n = 11$ (bottom figure) obtained with linear ORB5 simulations. While the thin white dots represents the Alfv\'en continuum computed by LIGKA, the cyan dots identifies the locations where the continuum is unstable. The radial coordinate $\rho_{t,orig}$ denotes the GENE radial grid while the coordinate $s = \sqrt{\Psi}$ denotes the ORB5 grid.}
\label{fig:spectr}
\end{figure}

Insights into the characterization of the energetic-particle-driven mode are provided with ORB5 and LIGKA~\cite{Lauber_JCP_2007}. Fig.~\ref{fig:spectr} shows the radial frequency spectrogram of the ORB5 linear simulations obtained for $n = 7$ (left) and $n = 11$ (right). To better identify this Alfv\'enic mode we add to Fig.~\ref{fig:spectr} the Alfv\'en continuum computed by LIGKA (we note that small but still finite differences are present in the q profiles used in the ORB5 and LIGKA simulations due to difficulties in matching CHEASE and HELENA~\cite{HELENA} for this case). We note that the normalization of the frequency spectrogram varies at each radial location, thus not allowing a direct comparison of intensity levels between different locations. Fig.~\ref{fig:spectr} shows that the high-frequency mode observed in the GENE and ORB5 linear simulations is destabilized at the Alfv\'en continuum (suggesting that the SAW continuum is not an effective damping mechanism for this mode) with a frequency that lies below the TAE gap. These modes are identified as a mixture of Alfv\'enic ion temperature gradient modes (AITGs) \cite{Zonca_1998,Hayward-Schneider_NF_2022} and kinetic beta-induced Alfv\'en eigenmodes (KBAEs) \cite{Heidbrink_PoP_1999}. It is worth mentioning that simulations at $n = 6$ without fast particles were performed showing an unstable narrow mode close to the $q= 8/6$ rational surface.

\section{Global electromagnetic GENE simulations} \label{sec5}

\begin{figure*}
\begin{center}
\includegraphics[scale=0.42]{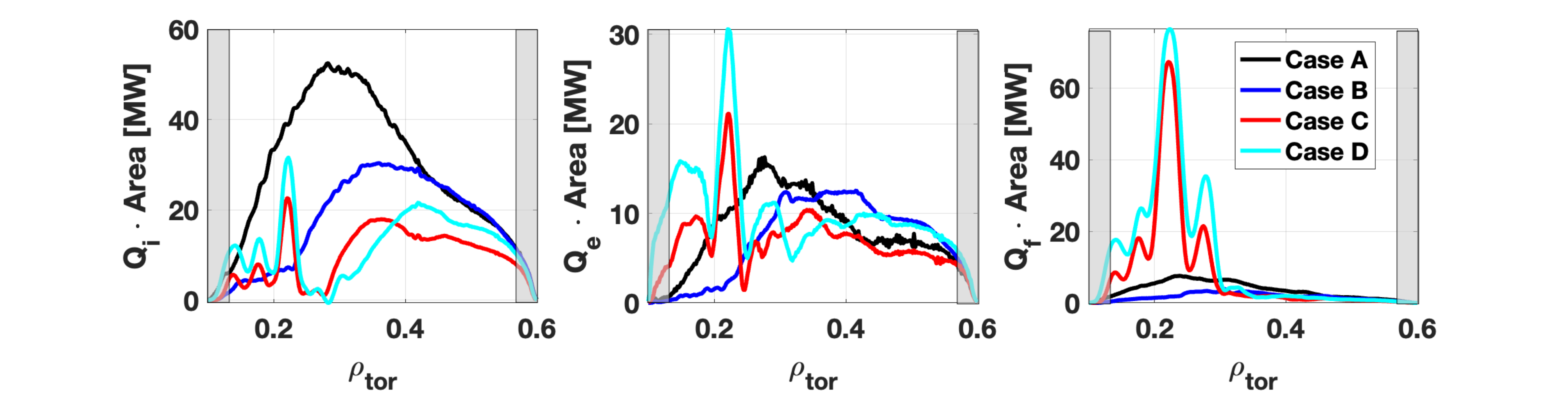}
\par\end{center}
\caption{Radial profile of the a) thermal ion, b) electron and c) supra-thermal particle heat fluxes obtained in the global GENE simulations over the saturated phase for the four different regimes of Fig.~\ref{fig:fig2}.}
\label{fig:fig_nl}
\end{figure*}

We discuss, in this Section, nonlinear global GENE turbulence simulations for the four different cases previously introduced. They only differ with respect to the plasma beta profiles (see, e.g., Fig.~\ref{fig:fig2} for $\beta_e$), which have been artificially modified in the underlying equations without changing the temperature and density profiles. This analysis allows us to investigate the impact of electromagnetic effects on the turbulence levels without affecting the pressure profiles that would inevitably impact e.g., plasma collisions. The numerical setup and resolutions employed for these global simulations are summarized in Tab.~\ref{table:tab1}. The time-averaged turbulent fluxes over the saturated nonlinear phase for each of the plasma species considered are shown in Fig.~\ref{fig:fig_nl}.
Fig.~\ref{fig:fig_nl} shows large turbulent fluxes in the electrostatic simulations (Case A) for each channel, with large values in the region where the logarithmic temperature gradients (for both thermal ions and electrons) increase, i.e., $\rho_{tor} = [0.2 - 0.4]$ (see Fig.~\ref{fig:fig1}). As electromagnetic effects at the nominal beta (Case B) are included in the simulations, we observe a significant (roughly $\sim 85\%$ at $\rho_{tor} \sim 0.2$ going from $\sim 33$MW to $\sim 5$MW) turbulence suppression for the thermal ion heat flux. For $\rho_{tor} > 0.4$, the turbulent levels recover the ones obtained in the electrostatic simulations. This is the radial domain where the energetic particle density is strongly reduced (going from $n_f / n_e \sim 0.11$ at $\rho_{tor} = 0.2$ to $n_f / n_e \sim 0.04$ at $\rho_{tor} = 0.42$), thus suggesting that the turbulence reduction in $\rho_{tor} = [0.2 - 0.4]$ is related to electromagnetic fast ion effects. This is consistent with previous flux-tube findings for the same plasma discharge \cite{Citrin_PRL_2013,DiSiena_NF_2019}. 

While the energetic particle fluxes undergo a similar reduction when including finite-$\beta$ effects, we notice that the electron heat flux is reduced in $\rho_{tor} = [0.1 - 0.3]$, but increases above the electrostatic ones for $\rho_{tor} > 0.3$. This is consistent with Ref.~\cite{DiSiena_JPP_2021}, showing that electromagnetic effects might act differently on the electron fluxes depending on the presence of fast ion effects. In particular, it was found that all the turbulence channels are stabilized in the region where the nonlinear electromagnetic fast ion effects are present. On the other hand, in the absence of fast particles, electromagnetic effects lead to a negligible thermal ion flux reduction, and to a destabilization of the electron heat flux, due to an increase of the electromagnetic electron flux (flutter) levels. This is in agreement with the results of Fig.~\ref{fig:fig_nl}. 

When the plasma beta is rigidly increased above its nominal values until destabilizing the AITG/KBAE (Case C), we observe from Fig.~\ref{fig:fig_nl} a strong increase of all turbulent fluxes in the radial domain where the AITG/KBAE is weakly destabilized, i.e., $\rho_{tor} = [0.1 - 0.3]$. While the ion heat flux increases by roughly a factor of 5 at $\rho_{tor} = 0.23$ (going from $\sim 5$MW to $\sim 23$MW), the electron flux increases by more than an order of magnitude (going from $\sim 2$MW to $\sim 21$MW) and the energetic particle flux by more than an order of magnitude (going from $\sim 1$MW to $\sim 70$MW). Moreover, a turbulence reduction is found in the ion heat flux compared to Case B for $\rho_{tor} > 0.3$ and spreads across the whole outer radial domain. This stabilization is likely caused by modifications of the turbulent heat flux radial propagation. In particular, the turbulent fluxes in these locations are significantly affected by the pronounced turbulent suppression found in the inner core locations. 

By further increasing the plasma beta (Case D), the drive of the AITG/KBAEs is enhanced with a consequent further destabilization of the energetic particle heat flux (see Fig.~\ref{fig:fig_nl}). The strongly unstable AITG/KBAEs lead to a significant increase in the electron and thermal ion heat fluxes in the whole radial domain where the AITG/KBAE is destabilized. Moreover, we also notice a mild destabilization of thermal ion and electron heat fluxes for $\rho_{tor} > 0.4$ compared to Case C.

\section{Comparison between flux-tube and global turbulent fluxes} \label{sec6}

A series of flux-tube (local) simulations are performed for each of the four different cases analyzed with global simulations to assess the range of validity of the local models in capturing energetic particle effects on turbulent transport. For each case, GENE (local) is run at five different locations, i.e., $\rho_{tor} = [0.18, 0.23, 0.28, 0.33, 0.43]$. 

\subsubsection{Case A: electrostatic limit}

\begin{figure*}
\begin{center}
\includegraphics[scale=0.42]{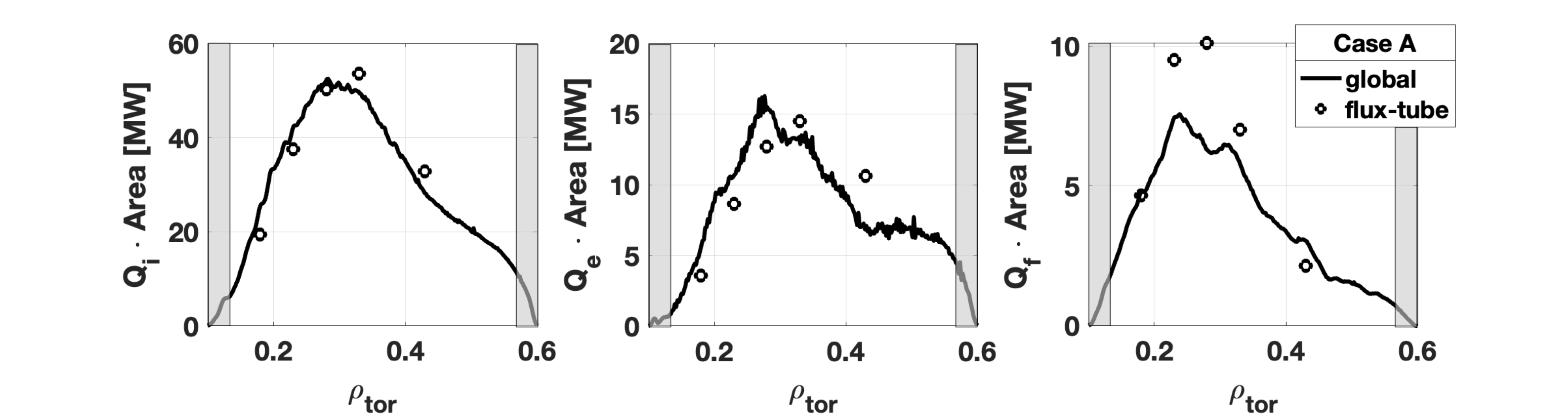}
\par\end{center}
\caption{Comparison of the radial profile of the a) thermal ion, b) electron and c) supra-thermal particle heat fluxes obtained in the global and flux-tube GENE simulations at four different locations for Case A (electrostatic, i.e., $\beta_e = 1e-4$).}
\label{fig:fig_nlA}
\end{figure*}

We begin by comparing the radially global and flux-tube (local) turbulent fluxes obtained for Case A (electrostatic limit). The results are illustrated in Fig.~\ref{fig:fig_nlA}. 
We find an excellent agreement between the flux-tube and global results for the thermal species (ions and electrons), with maximum deviations of $\sim 10\%$. The flux-tube simulations also show a quantitative good agreement with the global fast ion heat flux for $\rho_{tor} = 0.33$ and $0.43$. Larger differences are observed as we move towards the innermost locations (up to $\sim 30\%$), although still well reproducing the shape of the global fast ion flux.

\subsubsection{Case B: marginally stable AITG/KBAEs (nominal $\beta_e$)}

\begin{figure*}
\begin{center}
\includegraphics[scale=0.42]{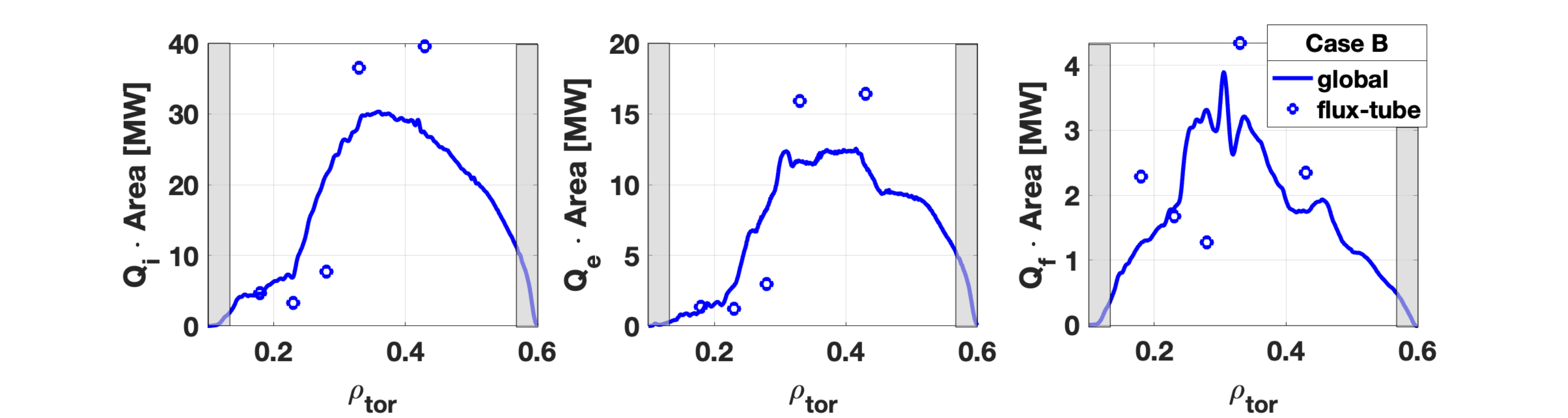}
\par\end{center}
\caption{Comparison of the radial profile of the a) thermal ion, b) electron and c) supra-thermal particle heat fluxes obtained in the global and flux-tube GENE simulations at four different locations for Case B (nominal).}
\label{fig:fig_nlB}
\end{figure*}

The heat flux comparison of Case B is particularly relevant since this is the plasma scenario corresponding to the nominal electron beta profile with marginally stable AITG/KBAEs. Under these conditions, flux-tube simulations show that - despite being linearly stable - AITG/KBAEs are destabilized nonlinearly depleting the energy content of the ITG turbulence, increasing the zonal flow levels, thus strongly suppressing ion-scale turbulence. Whether this mechanism is still relevant in radially global gyrokinetic simulations is still an open question. This is partially addressed in this section by comparing the turbulent fluxes of a radially global simulation with the ones obtained by running the flux-tube GENE version at the radial locations $\rho_{tor} = [0.18, 0.23, 0.28, 0.33, 0.43]$ for the scenario called Case B. The results are shown in Fig.~\ref{fig:fig_nlB}. This analysis will be further extended in the following section by looking at the frequency spectra, flux-surface averaged radial electric field and zonal currents to assess the limits of flux-tube simulations in modeling fast ion effects on turbulence. 
While the flux-tube simulations over-predicts the fast ion turbulence suppression by $\sim 40\%$ at $\rho_{tor} = [0.18, 0.23, 0.28]$, they predict larger turbulent fluxes by $\sim 30\%$ for $\rho_{tor} = [0.33, 0.43]$. Regardless of these differences, a good agreement is observed for the thermal species turbulent fluxes at each radial location. In particular, the flux-tube simulations capture well the sharp reduction of the thermal ion and electron fluxes at $\rho_{tor} < 0.33$. This is the radial domain where the energetic particle pressure is larger, and their impact on ion-scale turbulence is expected to be more effective. These observations are consistent with the experimental measurements showing a pronounced peaking of the ion temperature for $\rho_{tor} < 0.33$. We note that the trend of the fast ion turbulent flux around $\rho_{tor} = 0.23$ is not reproduced correctly in the flux-tube simulations.

\subsubsection{Case C: weakly unstable AITG/KBAEs}

\begin{figure*}
\begin{center}
\includegraphics[scale=0.42]{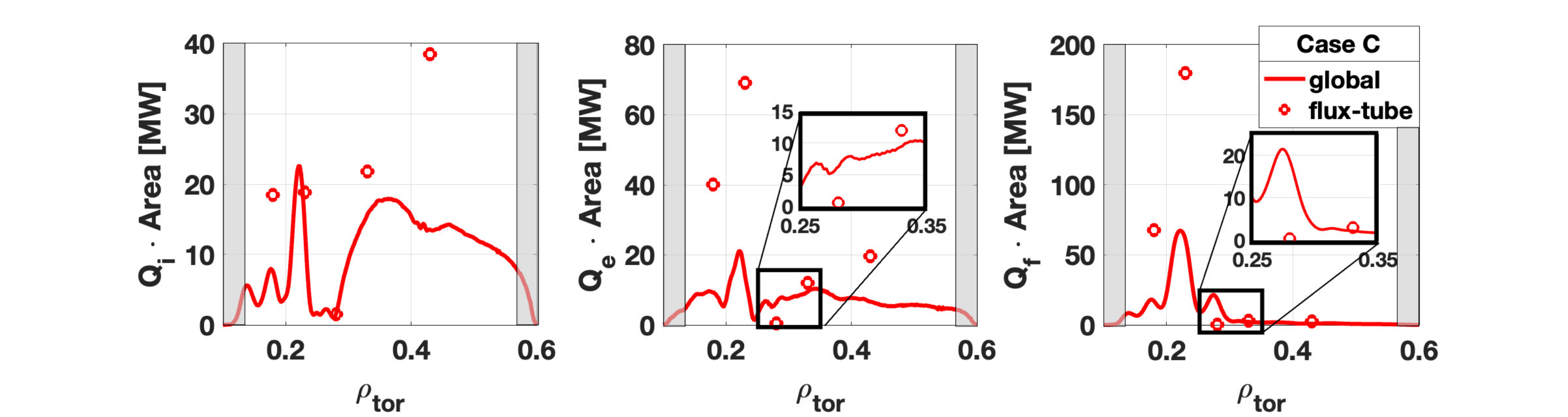}
\par\end{center}
\caption{Comparison of the radial profile of the a) thermal ion, b) electron and c) supra-thermal particle heat fluxes obtained in the global and flux-tube GENE simulations at four different locations for Case C (weakly unstable AITG/KBAEs). The inlays in Fig.~\ref{fig:fig_nlC}b-c show a zoom into the electron and fast ion turbulent fluxes in $\rho_{tor} = [0.25 - 0.35]$.}
\label{fig:fig_nlC}
\end{figure*}

The heat flux comparison between radially global and flux-tube results is performed here for Case C, i.e., when the plasma beta profile is increased until weakly destabilizing AITG/KBAEs. Similarly, as for the other cases studied above, five different radial locations have been selected, i.e., $\rho_{tor} = [0.18, 0.23, 0.28, 0.33, 0.43]$ for the local simulations. The results are illustrated in Fig.~\ref{fig:fig_nlC}.
A relatively good agreement between the flux-tube and global heat fluxes is observed - in each turbulent channel - only for the radial location $\rho_{tor} = 0.33$. While the flux-tube simulation at $\rho_{tor} = 0.43$ over-predicts by more than a factor of two the turbulent fluxes for both thermal ions and electrons, it predicts practically zero-fluxes at $\rho_{tor} = 0.28$. However, the far worst agreement is found at the radial location where the AITG/KBAE is located, i.e., $\rho_{tor} = 0.18$ and $\rho_{tor} = 0.23$. Here, the flux-tube simulation cannot reproduce the global turbulent fluxes for the electrons and the energetic particle species, which are heavily over-estimated by more than a factor of three. The large electron (mostly electromagnetic) and energetic particle (mostly electrostatic) heat fluxes observed in the flux-tube simulations are consistent with previous findings, showing that linearly unstable fast particle modes cannot be sustained in local simulations. This is due to the large over-prediction of the turbulent fluxes computed by the local simulations in these conditions, which are well above the experimental power balance. A global treatment is hence required in these scenarios to correctly model the radial mode structure of the AITG/KBAEs and the zonal patterns arising. This is shown in detail in the following sections. The flux-tube over-predicts the turbulent fluxes of the thermal ions by more than a factor of two at $\rho_{tor} = 0.18$.

\subsubsection{Case D: strongly unstable AITG/KBAEs}

\begin{figure*}
\begin{center}
\includegraphics[scale=0.42]{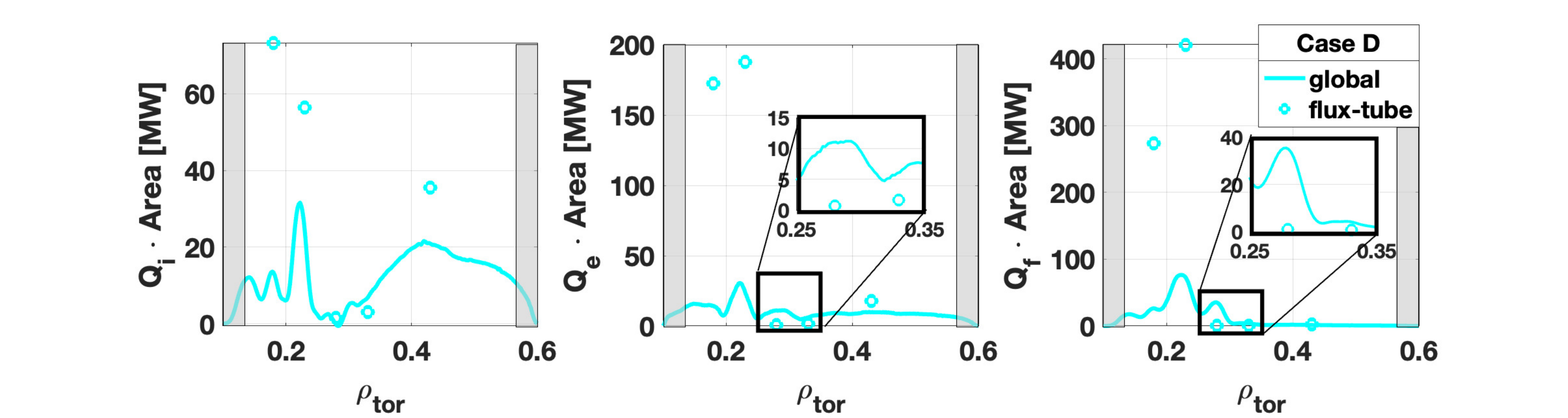}
\par\end{center}
\caption{Comparison of the radial profile of the a) thermal ion, b) electron and c) supra-thermal particle heat fluxes obtained in the global and flux-tube GENE simulations at four different locations for Case D (strongly unstable AITG/KBAEs). The inlays in Fig.~\ref{fig:fig_nlD}b-c show a zoom into the electron and fast ion turbulent fluxes in $\rho_{tor} = [0.25 - 0.35]$}
\label{fig:fig_nlD}
\end{figure*}

When the drive of the linearly unstable AITG/KBAE is further increased (Case D), the agreement between the flux-tube and radially global heat fluxes further degrades. This can be seen in Fig.~\ref{fig:fig_nlD}, where the global fluxes are compared with the ones computed with the local code at the five different locations $\rho_{tor} = [0.18, 0.23, 0.28, 0.33, 0.43]$.
While the local code largely overestimates the fluxes for each plasma species at $\rho_{tor} = 0.18$, $\rho_{tor} = 0.23$ and $\rho_{tor} = 0.43$, it underestimates the ones at $\rho_{tor} = 0.28$ and $\rho_{tor} = 0.33$. These findings are consistent with the previous results of Case C. The AITG/KBAE is strongly localized at the radial locations  $\rho_{tor} = 0.18$ and $\rho_{tor} = 0.23$, and it is stable at $\rho_{tor} > 0.28$ in the flux-tube simulations for all cases studied in this manuscript (not shown here). At the AITG/KBAE location $\rho_{tor} = 0.18$ and $\rho_{tor} = 0.23$, we find large differences between flux-tube and global results. Similarly, as for Case C, this indicates that the nonlinear AITG/KBAE nonlinear dynamics is not captured correctly in the local runs. This leads to different zonal flow levels and (less pronounced) zonal currents, thus affecting the local heat flux calculation at these locations. This is discussed in more detail in Section \ref{sec8}. On the other hand, in the radial domain $\rho_{tor} = [0.28 - 0.33]$, the increase in $\beta_e$ pushes the AITG/KBAEs closer to the marginal stability threshold. This leads to a particularly strong nonlinear turbulence suppression for the flux-tube simulations. This very localized stabilization is mitigated in the global GENE simulations due to the direct destabilization of the ion-scale turbulent fluxes by the AITG/KBAEs in the inner core regions. Finally, at $\rho_{tor} = 0.43$, the energetic particle effects are negligible in the flux-tube simulations due to the small fast ion pressure. This is consistent with the lack of sensitivity of the flux-tube turbulent fluxes to $\beta_e$ at this location. However, in the global simulations, we observe that the turbulent heat fluxes at $\rho_{tor} = 0.43$ are kept at low values by the significant turbulent suppression found in the inner core locations. This is not correctly captured in the flux-tube models, where the fluxes at each radial location are independent of the neighbouring positions.

\section{Frequency spectra in flux-tube and global simulations} \label{sec7}

The heat flux analyses summarized in the previous section show that a satisfying agreement between the flux-tube and global fluxes is achieved when the AITG/KBAEs are marginally stable. When the AITG/KBAEs are linearly unstable, the flux-tube simulations predict a strong increase in the turbulent fluxes at the AITG/KBAE location, mitigated in the global simulations. Therefore, at these positions ($\rho_{tor} \sim 0.18$ and $\rho_{tor} \sim 0.23$), the flux-tube runs significantly over-predict the turbulent fluxes for all plasma species.

\begin{figure*}
\begin{center}
\includegraphics[scale=0.35]{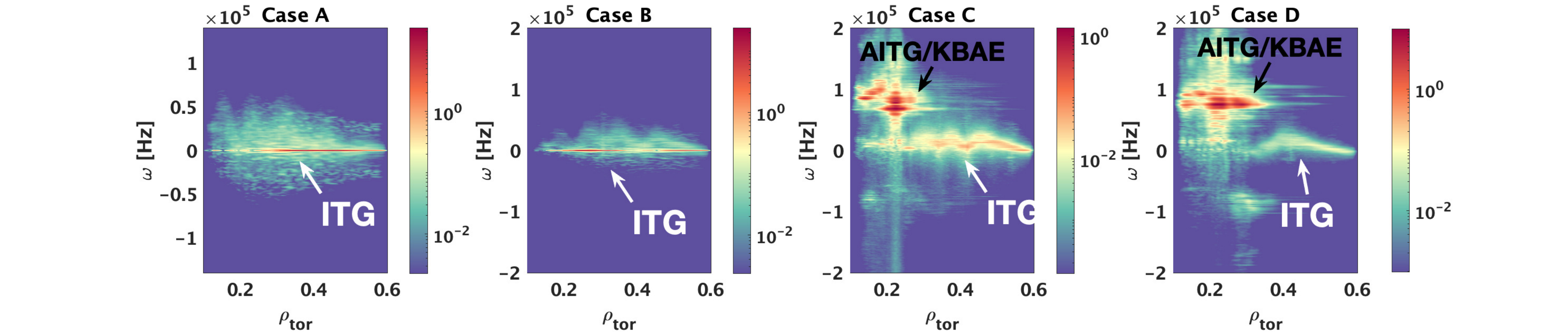}
\par\end{center}
\caption{Radial profile of the frequency spectra of the electrostatic potential $\phi_1$ obtained in the GENE global simulations in the time domain $t[c_s / a] = [200 - 450]$ for the four different regimes of Fig.~\ref{fig:fig2}. The electrostatic potential has been averaged over the toroidal mode numbers and field-aligned coordinate $z$. The amplitude of the signal is plotted in logarithmic scale. Note the different signal amplitude for Case C.}
\label{fig:fig_nl_fx}
\end{figure*}

In this section, we further expand our previous analyses by comparing the frequency spectra of the electrostatic potential for the flux-tube and global simulations. We begin by showing in Fig.~\ref{fig:fig_nl_fx} the frequency spectra profiles obtained in the global simulations for each case considered. The signals are averaged over the field-aligned coordinate.
Fig.~\ref{fig:fig_nl_fx} clearly shows a particularly broad structure at high-frequency for Case C and Case D localized in $\rho_{tor} = [0.15 - 0.3]$. This is the radial region where the turbulent fluxes strongly increase in the nonlinear global GENE simulations (see Fig.~\ref{fig:fig_nl}).

\begin{figure*}
\begin{center}
\includegraphics[scale=0.35]{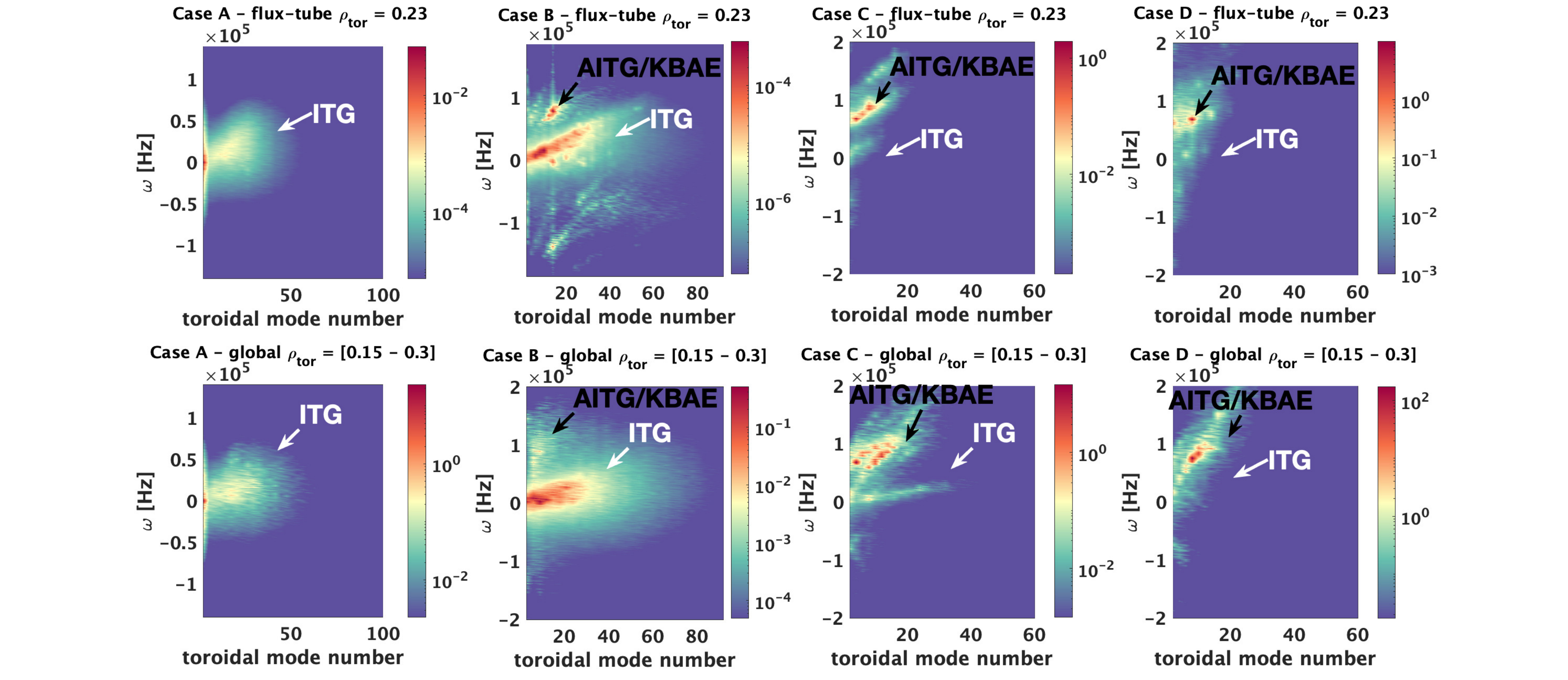}
\par\end{center}
\caption{Frequency spectra of the electrostatic potential $\phi_1$ obtained for the four different regimes of Fig.~\ref{fig:fig2} for the flux-tube (upper raw) at $\rho_{tor} = 0.23$ and global (lower raw) simulations in the time domain $t[c_s / a] = [200 - 450]$. While the flux-tube spectra are averaged over the field-aligned coordinate $z$ and the radial mode number $k_x \rho_s$, the radially global spectra on the field-aligned coordinate $z$ and the radial domain $\rho_{tor} = [0.15 - 0.3]$. The amplitude of the signal is plotted in logarithmic scale.}
\label{fig:fig_nl_f}
\end{figure*}

\begin{figure*}
\begin{center}
\includegraphics[scale=0.40]{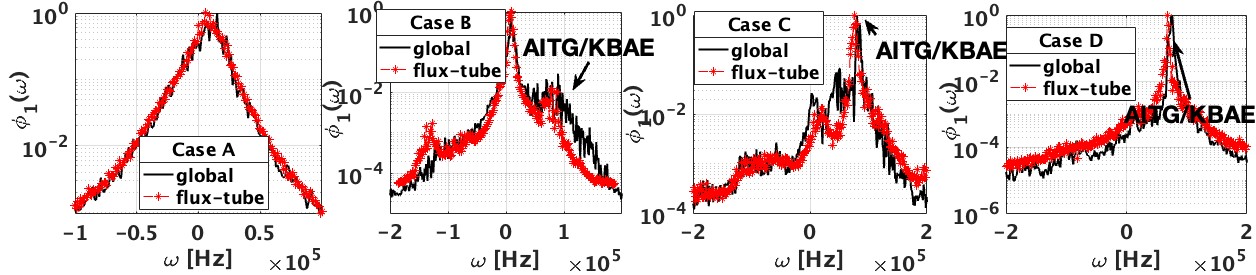}
\par\end{center}
\caption{Slices at $n = 6$ of Fig.~\ref{fig:fig_nl_f} showing a comparison of the frequency spectra of the electrostatic potential in the time domain $t[c_s / a] = [200 - 450]$ between flux-tube and global GENE simulations. While the flux-tube spectra are averaged over the field-aligned coordinate $z$ and the radial mode number $k_x \rho_s$, the radially global spectra on the field-aligned coordinate $z$ and the radial domain $\rho_{tor} = [0.15 - 0.3]$. The amplitude of the signal is plotted in logarithmic scale. To accentuate the distinctions between the local and global spectra, we employed a continuous black line for the global results and a red dotted line for the local ones for each case, that is labeled in each sub-figure legend.}
\label{fig:fig_nl_f_1d}
\end{figure*}

In Fig.~\ref{fig:fig_nl_f} we show the frequency spectra of the electrostatic potential for different toroidal mode numbers at each case of interest. The flux-tube spectra are shown only for the radial location $\rho_{tor} = 0.23$ which is the position where we observe the strongest turbulence suppression (Case B) and the linear destabilization of AITG/KBAEs (Case C and Case D). While the flux-tube spectra are averaged over the radial wave-numbers, the global spectra are averaged over the radial domain $\rho_{tor} = [0.15 - 0.3]$ (mainly to reduce numerical noise), thus including the location of the flux-tube runs. Both flux-tube and global spectra are averaged over the field-aligned coordinate.
For a closer comparison between the flux-tube and the global spectra, we show in Fig.~\ref{fig:fig_nl_f_1d}. a slice of the electrostatic potential at $n = 6$ (one of the most unstable toroidal mode numbers observed in the linear simulations of Fig.~\ref{fig:fig4} for the AITG/KBAE).

No qualitative differences are observed for the electrostatic potential spectra for Case A. However, when looking at Case B, we notice a more pronounced AITG/KBAE nonlinear destabilization in the flux-tube simulation compared to the global one. It is worth mentioning that the AITG/KBAEs are linearly stable for Case B but likely nonlinearly excited via cross-scale coupling with the ITG modes. The dynamics of this nonlinear interplay - observed in flux-tube simulations - was explained in Ref.~\cite{DiSiena_NF_2019}, and found to be essential to explain the strong ITG turbulence suppression in the presence of fast particles and electromagnetic effects. Interestingly, the global electrostatic potential spectra of Case B show clear signatures of unstable AITG/KBAEs, despite it being linearly stable. This finding is consistent with the significant turbulence suppression observed in the global GENE simulations of Case B with respect to the turbulent fluxes of Case A (where the fast ion electromagnetic stabilization is negligible due to the electrostatic setup), suggesting that the nonlinear coupling observed in flux-tube between ITGs and AITG/KBAEs take place similarly also in global simulations.

The results of Fig.~\ref{fig:fig_nl_f_1d} are also in agreement with the reduced turbulence suppression observed in the global simulation at $\rho_{tor} = [0.15 - 0.3]$ for Case B compared to the flux-tube results (see Fig.~\ref{fig:fig_nlB}). In particular, the flux-tube simulation exhibits a more pronounced nonlinear destabilization of AITG/KBAEs, thus further depleting the energy content of the ITGs and resulting in reduced turbulent fluxes. While this is also happening in the global simulation, the coupling between ITGs and AITG/KBAEs is partially mitigated by stronger damping of AITG/KBAEs at higher mode numbers (see Fig.~\ref{fig:fig_nl_f}).

When the AITG/KBAEs are linearly unstable (Case C and Case D), the electrostatic potential spectra of the flux-tube and global simulations show similar features (see Fig.~\ref{fig:fig_nl_f_1d}). In particular, the high-frequency AITG/KBAE branch overcomes the ITG one. This leads to a strong increase of the turbulent fluxes of all species in the radial domain where the AITG/KBAEs are located (i.e., $\rho_{tor} = [0.05 - 0.3]$, see Fig.~\ref{fig:fig_nl_f_1d}), although a reduction of the thermal ion fluxes is observed for $\rho_{tor} > 0.3$ (see Fig.~\ref{fig:fig_nl}.). This thermal ion heat flux reduction at $\rho_{tor} > 0.3$ is not captured in flux-tube simulations and is due to the excitation of radially global zonal flows and zonal current structures as discussed in the following sections.

\section{Flux-surface averaged radial electric fields and zonal currents} \label{sec8}

\begin{figure}
\begin{center}
\includegraphics[scale=0.31]{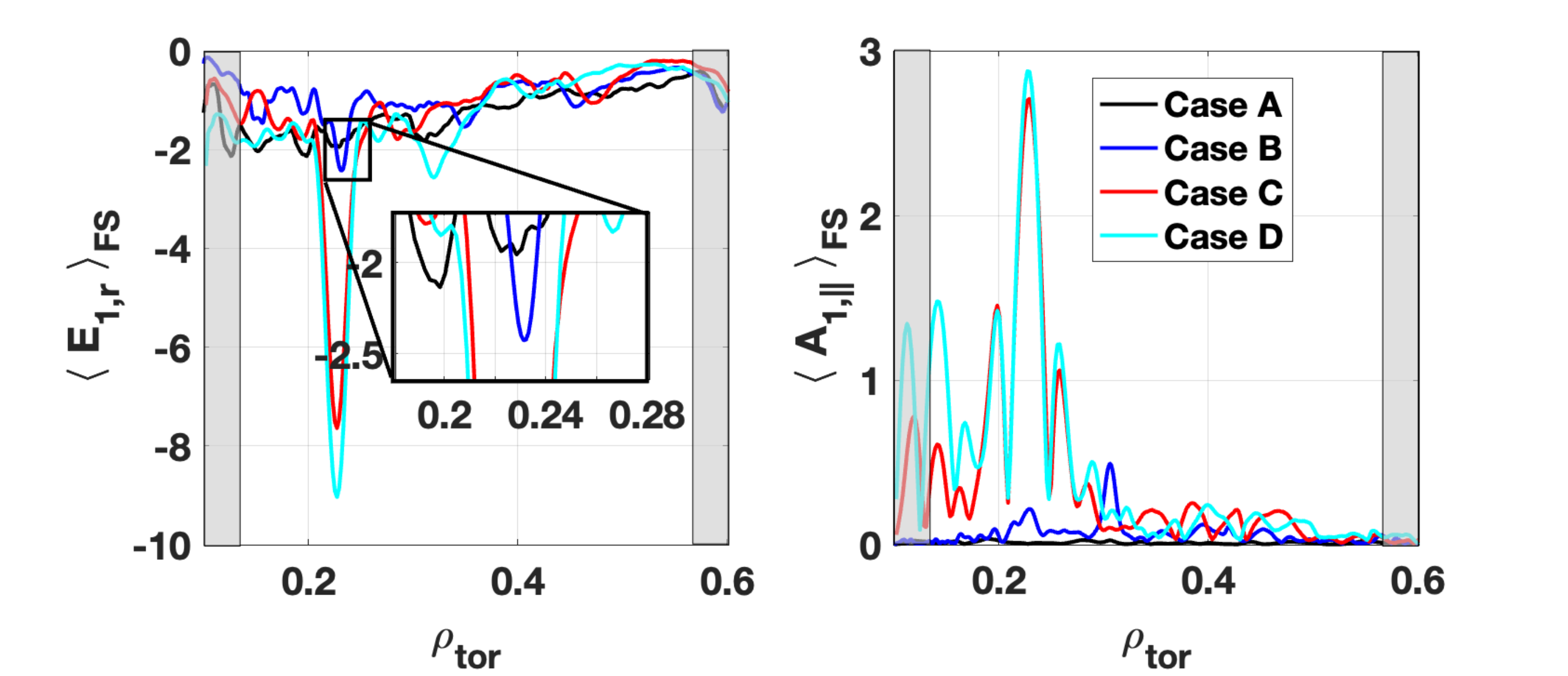}
\par\end{center}
\caption{Radial profile of the flux-surface averaged radial electric field $E_r = - \partial_{\rho_{tor}} \phi_1(n=0)$ (left plot) and zonal current $A_{1,\shortparallel}$ (right plot) obtained in the global GENE simulations over the saturated phase for the four different regimes of Fig.~\ref{fig:fig2}. The inlay in Fig.~\ref{fig:fig_ZF}a shows a zoom into the radial electric field in $\rho_{tor} = [0.18 - 0.28]$. The flux-surface averaged radial electric field and zonal current are normalized in GENE units \cite{Goerler_phd_thesis}.}
\label{fig:fig_ZF}
\end{figure}

\begin{figure*}
\begin{center}
\includegraphics[scale=0.40]{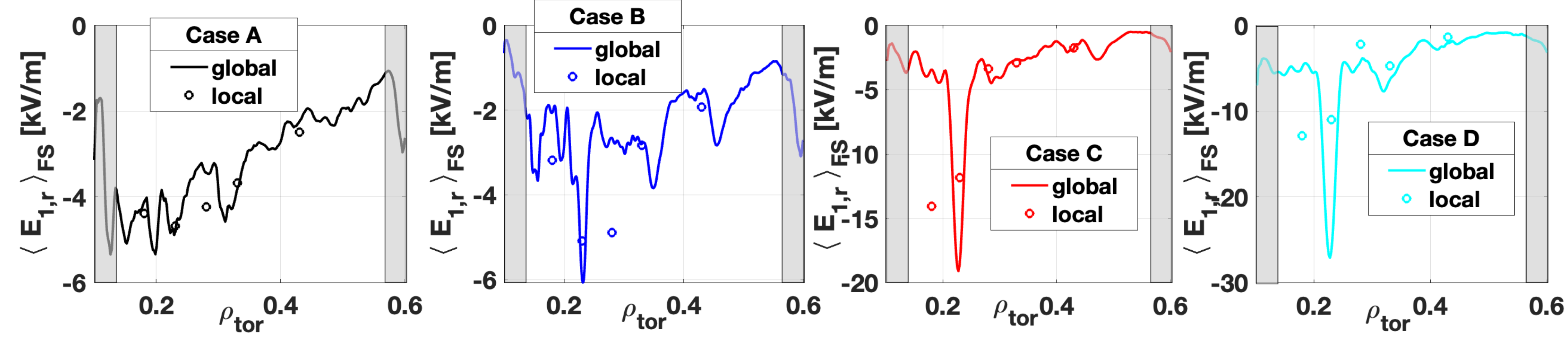}
\par\end{center}
\caption{Comparison of the flux-surface averaged radial electric field $E_r = - \partial_{\rho_{tor}} \phi_1(n=0)$ obtained in the flux-tube and global GENE simulations over the saturated phase for the four different regimes of Fig.~\ref{fig:fig2}. The flux-surface averaged radial electric field is expressed in SI units. Note the different signal amplitudes.}
\label{fig:fig_ZF_c}
\end{figure*}

Other observables compared in this manuscript to understand better the differences between the flux-tube and global results are the flux-surface averaged radial electric field and the zonal current. These quantities are one of the most effective drift-wave turbulence saturation mechanisms and can be driven by cross-scale coupling with the plasma micro-instabilities \cite{Diamond_PPCF_005} or induced by Alfv\'enic modes \cite{Zonca_PRL_2012}. We begin by showing in Fig.~\ref{fig:fig_ZF} the time-averaged radial profiles computed over the saturated nonlinear phase from the global simulations for each regime considered in this manuscript.
The flux-surface averaged radial electric field (defined as $E_r = - \partial_{\rho_{tor}} \phi_1(n=0)$) is de-localized through the whole radial domain for Case A. This can be seen more clearly in Fig.~\ref{fig:fig_ZF_c}a. A narrow structure appears for Case B at the location where the AITG/KBAEs are nonlinearly destabilized via cross-coupling with the ITG turbulence. This is shown more clearly in the inlay of Fig.~\ref{fig:fig_ZF}a and in Fig.~\ref{fig:fig_ZF_c}b. This observation is again consistent with the physical explanation proposed with flux-tube simulations where the nonlinearly excited AITG/KBAEs were shown to also contribute to a strong increase in the zonal flow levels, thus further suppressing ion-scale turbulence. When the AITG/KBAEs are linearly unstable (Case C and Case D), we notice a significant increase in the amplitude of the radial electric field at the location where the AITG/KBAEs are localized in the linear simulations, i.e., $\rho_{tor} \sim 0.23$. The excitation of a flux-surface averaged radial electric field due to Alfv\'en modes was also observed by other global electromagnetic codes in different scenarios \cite{Biancalani_PPCF_2021,Ishizawa_NF_2021}.

\begin{figure*}
\begin{center}
\includegraphics[scale=0.40]{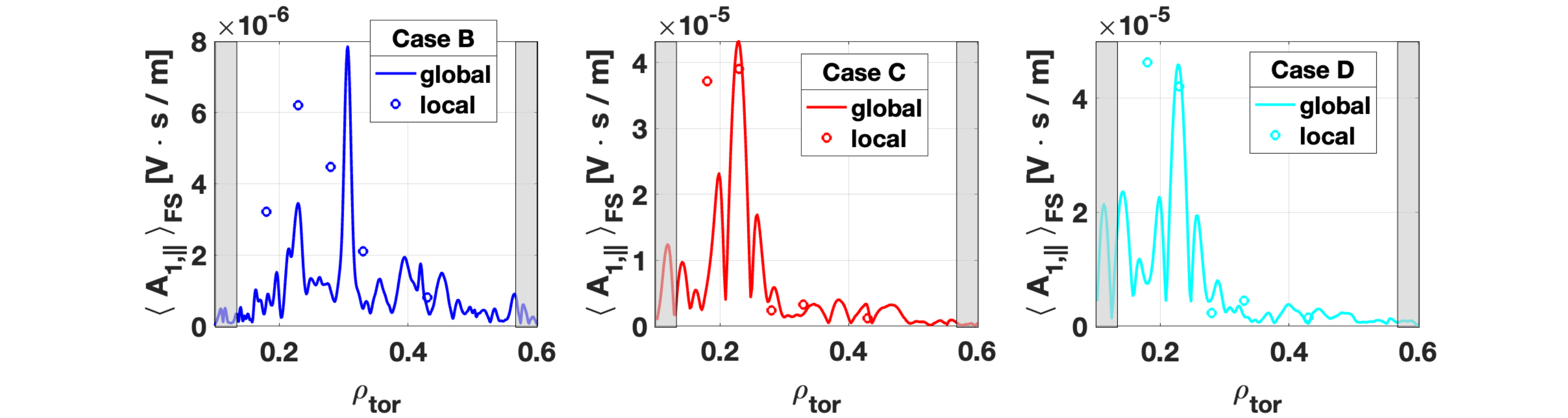}
\par\end{center}
\caption{Comparison of the flux-surface averaged zonal current $A_{1,\shortparallel}$ obtained in the flux-tube and global GENE simulations over the saturated phase for the four different regimes of Fig.~\ref{fig:fig2}. The flux-surface averaged zonal current is expressed in SI units. Note the different signal amplitudes.}
\label{fig:fig_A_c}
\end{figure*}

The flux-surface averaged zonal currents are illustrated in Fig.~\ref{fig:fig_ZF}b. While a localized increase is found for Case B at $\rho_{tor} \sim 0.23$ and $\rho_{tor} \sim 0.33$, we find that the AITG/KBAEs drive large zonal currents in the whole radial domain where their linear mode structure is located (see Fig.~\ref{fig:fig5} and Fig.~\ref{fig:fig_nl_fx}), i.e., $\rho_{tor} \sim [0.05 - 0.3]$. These findings show that while the AITG/KBAEs drive a localized flux-surface averaged radial electric field, they also lead to the excitation of radially broad zonal currents. This is well consistent with the theoretical predictions of Ref.~\cite{Zonca_PRL_2012}.
The global flux-surface averaged radial electric field and zonal currents are compared, respectively, in Fig.~\ref{fig:fig_ZF_c} and Fig.~\ref{fig:fig_A_c} with the ones obtained from the flux-tube simulations at the different locations for each case considered in this manuscript. 

An excellent agreement is observed for Case A for the radial electric field computed with the flux-tube and global simulations. When looking at Case B, we find a relatively good agreement with the only exception of the location $\rho_{tor} = 0.28$. At this position, the flux-tube simulation strongly over-predicts the amplitude of the radial electric field by roughly a factor of two. This is consistent with the heat flux analyses of Fig.~\ref{fig:fig_nlB}, showing the largest differences between the flux-tube and global heat fluxes at $\rho_{tor} = 0.28$. At this location, the flux-tube simulation under-predicts the turbulent fluxes for all turbulent channels. When the AITG/KBAE is linearly unstable (Case C and Case D), the flux-tube simulations fail in capturing the correct amplitude of the radial electric field at the location where the AITG/KBAE is destabilized (i.e., $\rho_{tor} = [0.18 - 0.23])$. The differences between flux-tube and radially global electric field increase at $\rho_{tor} = 0.23$ with the drive of the AITG/KBAE, while the radial electric field is underestimated by roughly a factor of two for Case C and three for Case D. These findings are again consistent with the heat flux analyses of Fig.~\ref{fig:fig_nlC} and Fig.~\ref{fig:fig_nlD}, where the global GENE simulations computed turbulent fluxes significantly lower than the corresponding flux-tube ones in the radial region where the AITG/KBAE was unstable. The radial electric field computed in flux-tube shows a relatively good agreement with the global simulations for $\rho_{tor} > 0.28$ for Case C and Case D.

Energetic-particle-driven modes can also induced zonal currents that can contribute to regulate drift-wave turbulence. In Fig.~\ref{fig:fig_A_c} we compare the zonal currents computed in the flux-tube and global simulations. Case A is not shown in this comparison given the nearly zero contribution of the zonal currents for the electrostatic setup $\beta_e \sim 1e-4$. Surprisingly, Case B shows the worst agreement with the flux-tube simulations over-predicting the amplitude of the zonal currents by more than a factor of two at $\rho_{tor} = 0.28$. In addition, the flux-tube runs do not recover the localized peak observed at $\rho_{tor} \sim 0.3$, which might explain the oscillatory pattern observed in the global fast ion heat flux of Fig.~\ref{fig:fig_nlB}c. This might be an indication that the global simulation for Case B captures some nonlinear interaction between the ITG turbulence and the linearly stable AITG/KBAEs which is not present in the flux-tube simulations. However, it is worth noticing that the relative amplitude of the zonal currents is relatively small for Case B and, hence, these differences might not impact the turbulent fluxes significantly. This is consistent with the relatively good agreement observed in the turbulent fluxes for Case B, despite the large differences of Fig.~\ref{fig:fig_A_c}. On the other hand, we find that the amplitude of the zonal currents computed in the flux-tube simulations well recover the global results for Case C and Case D with the only exception of $\rho_{tor} = 0.18$. This might be a further indication that the zonal currents do not contribute significantly to the turbulence saturation/regulation in these specific regimes.

\section{Conclusions} \label{sec9}

Most of the numerical analyses performed with gyrokinetic codes studying the effects of supra-thermal particles on core plasma turbulence have been performed using flux-tube codes, with only a few exceptions done on simplified setups \cite{Biancalani_PPCF_2021,Ishizawa_NF_2021}. This is due to the high complexity of running global gyrokinetic codes on these scenarios and to the enormous computational resources required for such global runs, requiring large radial, velocity and mode number resolutions. However, it is essential to assess whether flux-tube gyrokinetic simulations can correctly model energetic particle effects on core turbulence or if a radially global treatment is required. This is particularly important given the broad radial mode structure characterizing energetic-particle-driven modes and the large energetic particle temperatures making $1/\rho_f^* \sim [50 - 150]$ for most scenarios, potentially limiting the applicability of the numerically cheap and (typically) straightforward to run flux-tube simulations.

This manuscript focuses on this important issue presenting a detailed comparison between flux-tube and global simulations retaining supra-thermal particles. Our analyses are performed for realistic parameters inspired by a JET L-mode discharge showing a significant peaking of the ion temperature profile attributed to energetic particle effects on core plasma turbulence \cite{Mantica_PRL2011,Citrin_PRL_2013}. To extend our study to a broad range of different plasma scenario, this comparison is made by analyzing four different plasma regimes, which differ only by the plasma kinetic and magnetic pressure profile $\beta$. It has been artificially rescaled to cover the (i) electrostatic limit and regimes with (ii) marginally stable, (iii) weakly unstable and (iv) strongly unstable fast ion modes. This allows us to address the limits of flux-tube simulations in modeling energetic particle effects on core turbulence for each of these relevant scenarios.

We began our study by comparing the linear growth rates and frequencies of the energetic-particle-driven mode obtained from the global and flux-tube simulations. The latter were performed at the radial position where the mode structure of the energetic-particle-driven mode peaks for the most unstable toroidal mode number. A good qualitative agreement was observed between the linear flux-tube and global simulations with larger differences on the mode growth rates of $\sim 20\%$. The nature of this energetic-particle-driven mode was also identified as a AITG/KBAE via global ORB5 and LIGKA simulations after a successful benchmark with the global version of GENE.

The comparison between flux-tube and global GENE simulations was also extended to the nonlinear regime by comparing the heat flux levels for each plasma species selecting four different radial locations for the local simulations. We found that the flux-tube simulations reproduced qualitatively well the global results only in the electrostatic and marginally stable regimes, although the local runs overestimated the energetic particle turbulence stabilization by roughly $\sim 40\%$ for the marginally stable case. Despite this, the local results well reproduced the shape of the turbulent fluxes. These finding suggest the physical interpretation - primarily based on flux-tube simulations - involving marginally stable energetic particle modes suppressing ion-scale turbulent transport \cite{DiSiena_NF_2019,DiSiena_JPP_2021} might still hold in radially global setups. The local approximation failed to correctly model the turbulent fluxes in plasma regimes with the energetic-particle-driven mode linearly unstable. While the flux-tube simulations largely overestimated the turbulent fluxes (mainly for electrons and fast ions) in the radial domain where the AITG/KBAE is located ($\rho_{tor} = [0.18 - 0.23]$), they underestimated the fluxes at the neighbouring locations. The differences between flux-tube and global results increase with the linear drive of the AITG/KBAE.

These findings are explained in terms of a strong mismatch between the flux-tube and global simulations for the zonal flow levels where the energetic particle drive is destabilized. In particular, when comparing the flux-surface averaged radial electric field measured in the flux-tube simulations, we found that the local simulations failed in reproducing the global zonal patterns arising via the interaction between the AITG/KBAE and the zonal flows in the radial domain where the AITG/KBAE is located. This leads to a different turbulence regulation of the flux-tube simulations thus resulting in different turbulent fluxes. At finite toroidal mode numbers, no significant differences are observed in the electrostatic potential spectra between flux-tube and global simulations.

Therefore, our study suggest that global turbulence simulations are likely required in regimes with linearly unstable fast ion-driven modes. Techniques to identify the parameter space where such fast ion modes are unstable without running expensive simulations are currently under investigations.

\section*{Acknowledgements}

The authors would like to acknowledge insightful discussions with P. Lauber and F. Zonca. This work has been carried out within the framework of the EUROfusion Consortium, funded by the European Union via the Euratom Research and Training Programme (Grant Agreement No 101052200 — EUROfusion). Views and opinions expressed are however those of the author(s) only and do not necessarily reflect those of the European Union or the European Commission. Neither the European Union nor the European Commission can be held responsible for them. Numerical simulations were performed at the Marconi and Marconi100 Fusion supercomputers at CINECA, Italy.

\end{document}